\documentclass[]{aastex631}

\usepackage{graphicx}
\usepackage{amsmath}

\usepackage{multirow}

\def\code#1{\texttt{#1}}

\def\HeI{\hbox{He~$\scriptstyle\rm I$~}}
\def\HeII{\hbox{He~$\scriptstyle\rm II$~}}
\def\NII{\hbox{N~$\scriptstyle\rm II$~}}
\def\fNII{\hbox{[N~$\scriptstyle\rm II$]~}}
\def\fXeIII{\hbox{[Xe~$\scriptstyle\rm III$]~}}

\def\OII{\hbox{O~$\scriptstyle\rm II$~}}

\def\fOII{\hbox{[O~$\scriptstyle\rm II$]~}}
\def\fSII{\hbox{[S~$\scriptstyle\rm II$]~}}
\def\fSIII{\hbox{[S~$\scriptstyle\rm III$]~}}
\def\fArIII{\hbox{[Ar~$\scriptstyle\rm III$]~}}
\def\fArIV{\hbox{[Ar~$\scriptstyle\rm IV$]~}}
\def\fKrIII{\hbox{[Kr~$\scriptstyle\rm III$]~}}
\def\fKrIV{\hbox{[Kr~$\scriptstyle\rm IV$]~}}
\def\fKIV{\hbox{[K~$\scriptstyle\rm IV$]~}}

\def\fOIII{\hbox{[O~$\scriptstyle\rm III$]~}}
\def\fClII{\hbox{[Cl~$\scriptstyle\rm II$]~}}
\def\fClIII{\hbox{[Cl~$\scriptstyle\rm III$]~}}
\def\fClIV{\hbox{[Cl~$\scriptstyle\rm IV$]~}}
\def\fNeIII{\hbox{[Ne~$\scriptstyle\rm III$]~}}
\def\fArV{\hbox{[Ar~$\scriptstyle\rm V$]~}}
\def\fFeII{\hbox{[Fe~$\scriptstyle\rm II$]~}}
\def\fFeIII{\hbox{[Fe~$\scriptstyle\rm III$]~}}
\def\HI{\hbox{H~$\scriptstyle\rm I$~}}
\def\Gaia{\hbox{\it{Gaia}}}

\shorttitle{Abundances of Eight Planetary Nebulae with LRS2}
\shortauthors{Manea et al.}

\graphicspath{{./}{figures/}}

\begin{document}
\title{Chemical Abundances of Eight Highly Extincted Milky Way Planetary Nebulae\footnote{Based on observations obtained with the Hobby-Eberly Telescope, which is a joint project of the University of Texas at Austin, the Pennsylvania State University, Ludwig-Maximilians-Universit\"{a}t M\"{u}nchen, and Georg-August-Universit\"{a}t G\"{o}ttingen.}}

\author[0000-0002-0900-6076]{Catherine Manea}
\affiliation{University of Texas at Austin, Austin, TX, 78712, USA}

\author{Harriet L. Dinerstein}
\affiliation{University of Texas at Austin, Austin, TX, 78712, USA}

\author{N. C. Sterling}
\affiliation{University of West Georgia, Carrollton GA 30118, USA}

\author{Greg Zeimann}
\affiliation{Hobby-Eberly Telescope, University of Texas at Austin, Austin, TX, 78712, USA}

\begin{abstract}
Low- and intermediate-mass ($\rm 0.8~M_\odot < M < 8~M_\odot$) stars that evolve into planetary nebulae (PNe) play an important role in tracing and driving Galactic chemical evolution. Spectroscopy of PNe enables access to both the initial composition of their progenitor stars and products of their internal nucleosynthesis, but determining accurate ionic and elemental abundances of PNe requires high-quality optical spectra. We obtained new optical spectra of eight highly-extincted PNe with limited optical data in the literature using the Low Resolution Spectrograph 2 (LRS2) on the Hobby-Eberly Telescope (HET). Extinction coefficients, electron temperatures and densities, and ionic and elemental abundances of up to 11 elements (He, N, O, Ne, S, Cl, Ar, K, Fe, Kr, and Xe) are determined for each object in our sample. Where available, astrometric data from \Gaia{} eDR3 is used to kinematically characterize the probability that each object belongs to the Milky Way’s thin disk, thick disk, or halo. Four of the PNe show kinematic and chemical signs of thin disk membership, while two may be members of the thick disk. The remaining two targets lack \Gaia{} data, but their solar O, Ar, and Cl abundances suggest thin disk membership. Additionally, we report the detection of broad emission features from the central star of M 3-35. Our results significantly improve the available information on the nebular parameters and chemical compositions of these objects, which can inform future analyses.
\end{abstract}


\section{Introduction} \label{sec:intro}
Stars are chemical time capsules, trapping their initial elemental abundances at formation, upon which are superposed the effects of internal nuclear processing during their evolution. Planetary nebulae (PNe), composed of material that represents the final chemical composition at the end of the full nucleosynthetic histories of relatively long-lived low- and intermediate-mass ($\rm 0.8~M_\odot < M < 8~M_\odot$) stars, can be used to trace the past composition of the interstellar medium (ISM) as well as the net contributions of these stars to the chemical evolution of galaxies. Abundances in PNe of elements that are unaffected by internal nucleosynthesis such as O, Ar, S, and Cl are indicators of the parent star’s original composition and reflect its membership of a particular Milky Way population, such as the thin and thick disks, bulge, and halo. On the other hand, during the asymptotic giant branch (AGB) phase, stars that create PNe produce significant amounts of C, N, F, and Na as well as trans-iron nuclides synthesized by neutron captures in the s-process \citep[][]{Herwig2005, Kappeler2011, Karakas2014, Karakas2016, Amayo2020}.  Recent observations have detected optical, infrared, and even UV lines in PNe from elements such as Ge, Se, Kr, Rb, Te, and Xe despite their low abundances relative to H even in highly enriched objects \citep[][]{Pequignot1994, Dinerstein2001, Sterling2003, Sharpee2007, Sterling2008, Sterling2016, Madonna2018}.

Determinations of accurate elemental abundances in PNe rely on using diagnostic ratios that indicate physical conditions such as electron temperature and density. Since some of these diagnostics require measurements of weak lines, high-quality optical spectra are essential for such studies (e.g., see reviews of \citealt{Peimbert2017} and \citealt{Kwitter2022}). The optical spectral region also features lines of abundant elements such as He, O, Ar, etc. that are used to determine ionization correction factors (ICFs) that account for unseen ions of other elements that may contain significant fractions of their respective elements. These factors enable calculation of total elemental abundances when few ions (or only one ion) of a given element are observed, and most available ICF formulations utilize commonly observed optical emission lines \citep[e.g.,][hereafter SPD15]{Kingsburgh1994, Delgado2014, Sterling2015}.

In this work we present new optical spectra of eight PNe that lack previous high-quality data suitable for the determination of accurate elemental abundances, likely due to the high extinctions (2.1~\textless~c(H$\beta$)~\textless~4.2) of all but M 4-18. The sample was selected in part due to potential indications of s-process enrichments in their infrared spectra (SPD15, \citealp{Dinerstein2022} and in preparation). In Section \ref{sec:observations}, we describe our observations and the data reduction process. In Section \ref{sec:chemanalysis}, we discuss our methods for determining the physical conditions and ionic and elemental abundances of our sample. Section \ref{sec:chemresults} presents the results of our chemical analysis. In Section \ref{sec:kin}, we discuss the Galactic XYZ positions and UVW space velocities of each target and consider their probabilities of membership to the Milky Way thin disk, thick disk, and halo. We conclude with a summary in Section \ref{sec:sum}.

\section{Observations and Data Reduction} \label{sec:observations}
The targets in our sample were observed in queue mode \citep{Shetrone2007} across two trimesters using the Low Resolution Spectrograph 2 (LRS2, \citealt{LRS2}) on the upgraded Hobby-Eberly Telescope (HET, \citealt{Ramsey1998, Hill2021}) in west Texas. LRS2 is a low-resolution (R$\sim$1900) optical integral-field unit (IFU) spectrograph composed of two arms that simultaneously observe two 6\arcsec $\times$12\arcsec~fields of view separated by 100\arcsec. The blue arm consists of a pair of channels with spectral ranges of $\sim$3640--4670 $\rm\AA$ and $\sim$4540--7000 $\rm\AA$, while the red arm is composed of two channels covering $\sim$6430--8450 $\rm\AA$ and $\sim$8230--10560 $\rm\AA$. LRS2 is mounted on the prime focus tracker of the fixed altitude HET, and objects are observed in tracks with the maximum duration limited by the object’s coordinates. For each target, we obtained a short exposure in addition to a set of longer exposures in order to sample the full dynamic range of fluxes in the spectrum, with the short exposures avoiding saturation on the strongest lines and the longer exposures providing sufficient sensitivity to reach weaker lines. We used exposures from the arm not actively observing the target for the purpose of sky subtraction. Observations of standard stars were used for relative spectrophotometric flux calibration and telluric corrections. Table \ref{tab:targs} summarizes our observations.
\renewcommand{\arraystretch}{1.0}
\begin{deluxetable*}{llcccccl}
\tablenum{1}
\tablecaption{Targets and Observations \label{tab:targs}}
\tablewidth{0pt}
\tablehead{ \colhead{Target Name} & \colhead{PNG} & \colhead{Diameter\tablenotemark{\footnotesize a}} & \colhead{CS Type\tablenotemark{\footnotesize b}} & \colhead{Obs. Date} &  \multicolumn{2}{c}{Exposure Time} & \colhead{Observing Notes} \\
\colhead{ } & \colhead{ } & \colhead{(\arcsec)} & \colhead{ (mag) }& \colhead{(UT)}  & \colhead{Long (s)} & \colhead{Short (s)} & \colhead{ } 
}
\startdata
Hen 2-459 & 068.3-02.7 & 3.0 & [WC9] & 2020 Aug 07 & 480 & 30 & Clear \\
K 3-17 & 039.8+02.1 & 18.6 & H-rich & 2020 Sep 22 & 2100 & 120 &  Clear \\
K 3-55 & 069.7+00.0 & 9.0 & H-rich & 2020 Nov 20 & 1900 & 120 &  Clear \\
K 3-60 & 098.2+04.9 & 3.0 & H-rich & 2020 May 29 & 2674 & 90 &  Thin Clouds \\
K 3-62 & 095.2+00.7 & 5.0 & H-rich  & 2020 May 15 & 640 & 30 &  Thin Clouds \\
M 2-43 & 027.6+04.2 & 2.0  & [WC7-8] & 2020 Apr 18 & 370 & 30 &  Clear \\
M 3-35 & 071.6-02.3 & 4.6 & [WC]/WELS\tablenotemark{\footnotesize c} & 2020 Oct 23 & 480 & 30 &  Clear \\
M 4-18 & 146.7+07.6 & 3.7 & [WC11] & 2020 Oct 07 & 480 & 30 &  Clear \\
\enddata
\tablenotetext{\footnotesize a}{Optical diameters from the HASH PN database \citep{HASH}.  Diameters for non-circular nebulae are reported as averages of the minor and major axes.}
\tablenotetext{\footnotesize b}{Central star type from \citet{Acker2003} with the exception of M 3-35.}
\tablenotetext{\footnotesize c}{Although not previously reported in the literature, M 3-35 displays weak stellar wind lines (see Figure \ref{fig:spectra} and Section \ref{sec:emission-line}).}
\end{deluxetable*}

The raw data were processed with Panacea\footnote{\code{https://github.com/grzeimann/Panacea}}, an automated reduction pipeline for LRS2 written by G. Zeimann  (Zeimann et al. 2023, in preparation). Panacea performs an optimal extraction by fitting a 2-D Gaussian to all spaxels that contain high signal-to-noise emission features, extracting the data with weighting from the Gaussian model. Prior to extraction, the data are bias-corrected, flat-field-corrected, and sky-subtracted.  Panacea uses a response curve generated from standard stars over many nights to perform relative flux calibration.  After optimal extraction, the data are collapsed into 1-D.  The 1-D spectra from each arm of LRS2 are then stitched together using overlapping spectral regions so as to preserve relative fluxes.  The default sky-subtraction for Panacea uses the fibers within the active observation, but since our program observed LRS2-B and LRS2-R consecutively, we used the sky observations from the off-target exposure to avoid self-subtraction of real nebular lines.  The sky from the off-target exposure was scaled manually to match the sky in the science exposure to account for the non-simultaneous sampling. The final data products used in our analysis consist of two sets of fully-reduced 1-D spectra for each object, one each from the short and long exposures. Figure \ref{fig:spectra} displays representative excerpts from our spectra.

\section{Temperatures, Densities, and Chemical Abundances} \label{sec:chemanalysis}
\subsection{Line Fluxes} \label{subsec:measuringspectra}
To determine line fluxes and associated uncertainties, we first input the spectra into the Automated Line Fitting Algorithm, ALFA\footnote{\code{https://nebulousresearch.org/codes/alfa/}} (\citealp{ALFA}), which efficiently measures line fluxes in spectra of arbitrary wavelength range and resolution. ALFA performs a spline fit to the continuum using a 100 data point window, which is subtracted from the spectrum prior to emission line measurement. The lines are modelled as Gaussians with central wavelengths taken from a line list and modified to produce an array of lines that best fits the input spectrum. ALFA has an extensive line list constructed from deep optical spectra of Galactic PNe by \citet[][]{Wesson2005}. It also has deblending capabilities, which are important at the low resolution of LRS2.

\begin{figure}
    \centering
    \includegraphics[width=0.6\linewidth]{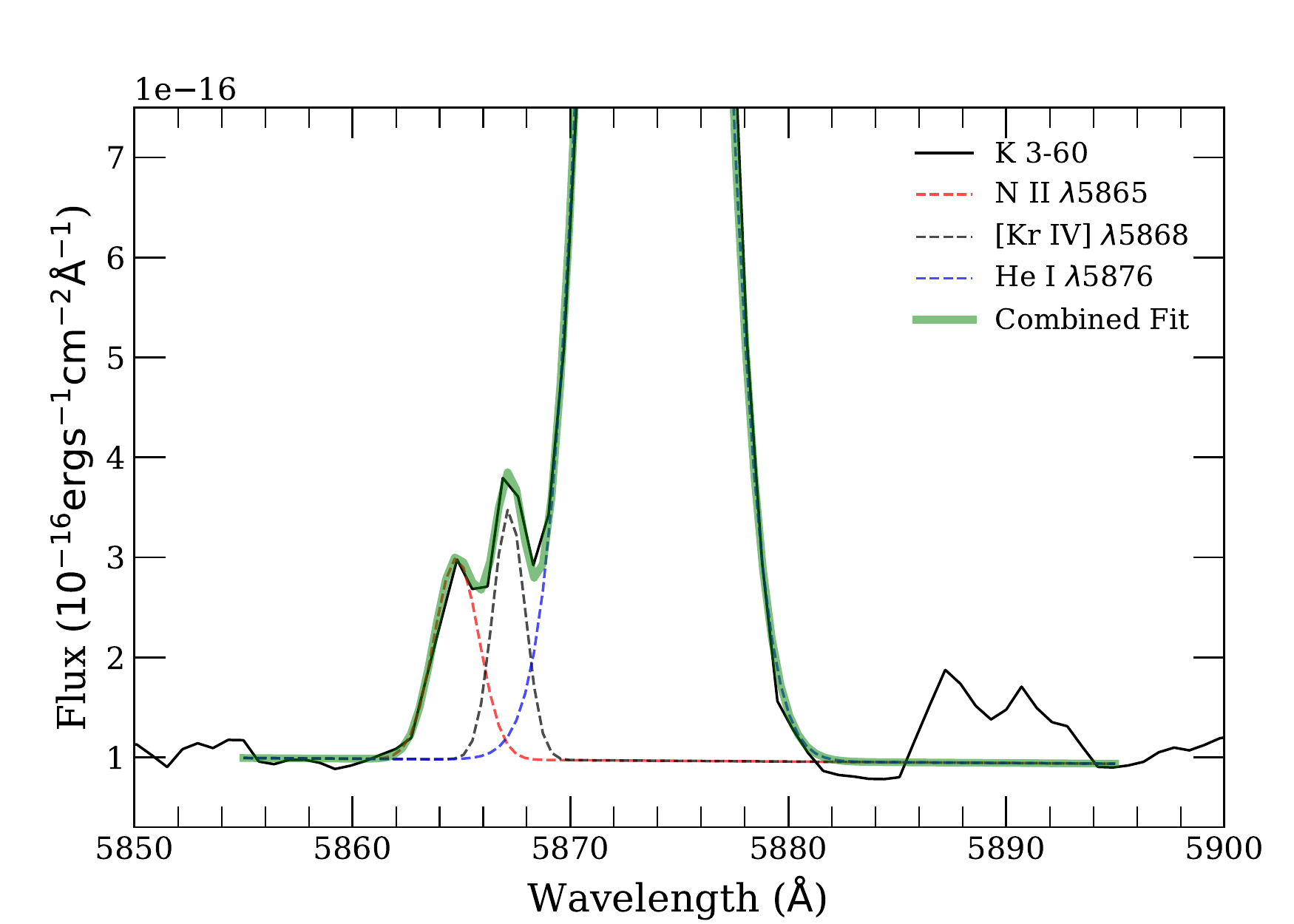}
    \caption{Example of the deblending of \NII $\lambda$5865 and \fKrIV $\lambda$5868, on the wing of the very strong \HeI $\lambda$5876 line, using IRAF's \code{splot} routine. The Na I D lines near 5890$\rm \AA$ are not included in the fit.  This spectral region lies in the orange arm of LRS2, which has a resolution of $\delta\lambda \sim 3 \rm \AA$.}
    \label{fig:iraf}
\end{figure}

We visually inspected the output from ALFA to monitor the quality of fits to the continuum and detected lines. ALFA was most successful at fitting strong and well-resolved features but had difficulty with some of the weaker and more highly blended lines, particularly where a weak line was blended with a much stronger one or was superposed on a broad stellar emission feature. In such cases, we used the IRAF task \code{splot}, estimating flux uncertainties by determining line flux values for three placements of the continuum: the optimal value and reasonable maximum and minimum levels. Figure \ref{fig:iraf} presents an example of an IRAF fit to a blend of lines of disparate strengths. We present the observed fluxes and associated uncertainties of lines used in our analysis in Table \ref{tab:fluxlist}. 

\subsection{Final Line Intensities} \label{subsec:lineints}
The observed relative emission line fluxes must be corrected for interstellar dust extinction. To determine the extinction of each nebula, we compare the observed flux ratio of H$\alpha$ to H$\beta$ from the short-exposure spectra to the theoretical ratio \citep[][]{Storey1995}. Before doing this comparison, however, we correct the H$\beta$ and H$\alpha$ fluxes for contributions from \HeII $\lambda$4859 and $\lambda$6560, respectively, which are blended with the \HI lines at the resolution of our data. We scale the blended \HeII line fluxes from the strong, isolated \HeII $\lambda$4686 line, where \HeII $\lambda$4859 $\cong$ 6\% $\times$ $\lambda$4686 and \HeII $\lambda$6560 $\cong$ 13\% $\times$ $\lambda$4686 \citep[][]{Osterbrock2006}. Assuming an intrinsic H$\alpha$ to H$\beta$ ratio of 2.86, we determine c(H$\beta$), the logarithmic extinction coefficient at H$\beta$, as reported in Table \ref{tab:tempdens}. To calculate line intensities corrected for extinction, we adopt the extinction law of \citet{Fitzpatrick1999}. 

Next, we apply corrections to certain lines in order to take additional line blends into account. In objects that display \HeII lines, the auroral \fSIII $\lambda$6312 line, which is a key component of an electron temperature diagnostic, is blended with \HeII $\lambda$6311 at LRS2 resolution. We subtract the expected intensity of \HeII $\lambda$6311 (0.36\% of \HeII $\lambda$4686) from the blend at $\lambda$6312 before using the \fSIII diagnostic. The \HeII contribution is most significant in K 3-17, where \HeII $\lambda$6311 comprises 8.8\% of the flux of \fSIII $\lambda$6312. In M 2-43 and M 3-35, permitted lines of \NII and \OII were detected, allowing us to  estimate the recombination contribution to \fNII $\lambda$5755 and the \fOII $\lambda$7325 quartet using the  prescriptions of \citet{Liu2000}. We determined permitted line abundances of N$^{++}$/H$^{+}$ and  O$^{++}$/H$^{+}$ from \NII multiplet 3 and \OII multiplets 1 and 2 respectively in M 3-35, while in M  2-43 only \NII $\lambda$5666 was sufficiently isolated from stellar features to be used in the analysis.  The recombination contribution to \fNII $\lambda$5755 in M 2-43 is negligible ($\sim$1\%). In M 3-35, we  find a recombination contribution of 6\% to the \fNII $\lambda$5755 intensity and approximately 10\% for \fOII $\lambda$7325. These corrections were applied in our abundance analysis. For the other targets, permitted lines of N~$\scriptstyle\rm II$, O~$\scriptstyle\rm II$, and O~$\scriptstyle\rm III$ were not detected due to high extinction and/or strong stellar features overwhelming the weak permitted lines, and hence recombination corrections could not be applied to [N~$\scriptstyle\rm II$], [O~$\scriptstyle\rm II$], and [O~$\scriptstyle\rm III$] lines.  We present the intensities and associated uncertainties of lines used in our analysis in Table \ref{tab:linelist}.

\begin{figure}
    \centering
    \includegraphics[width=1\linewidth]{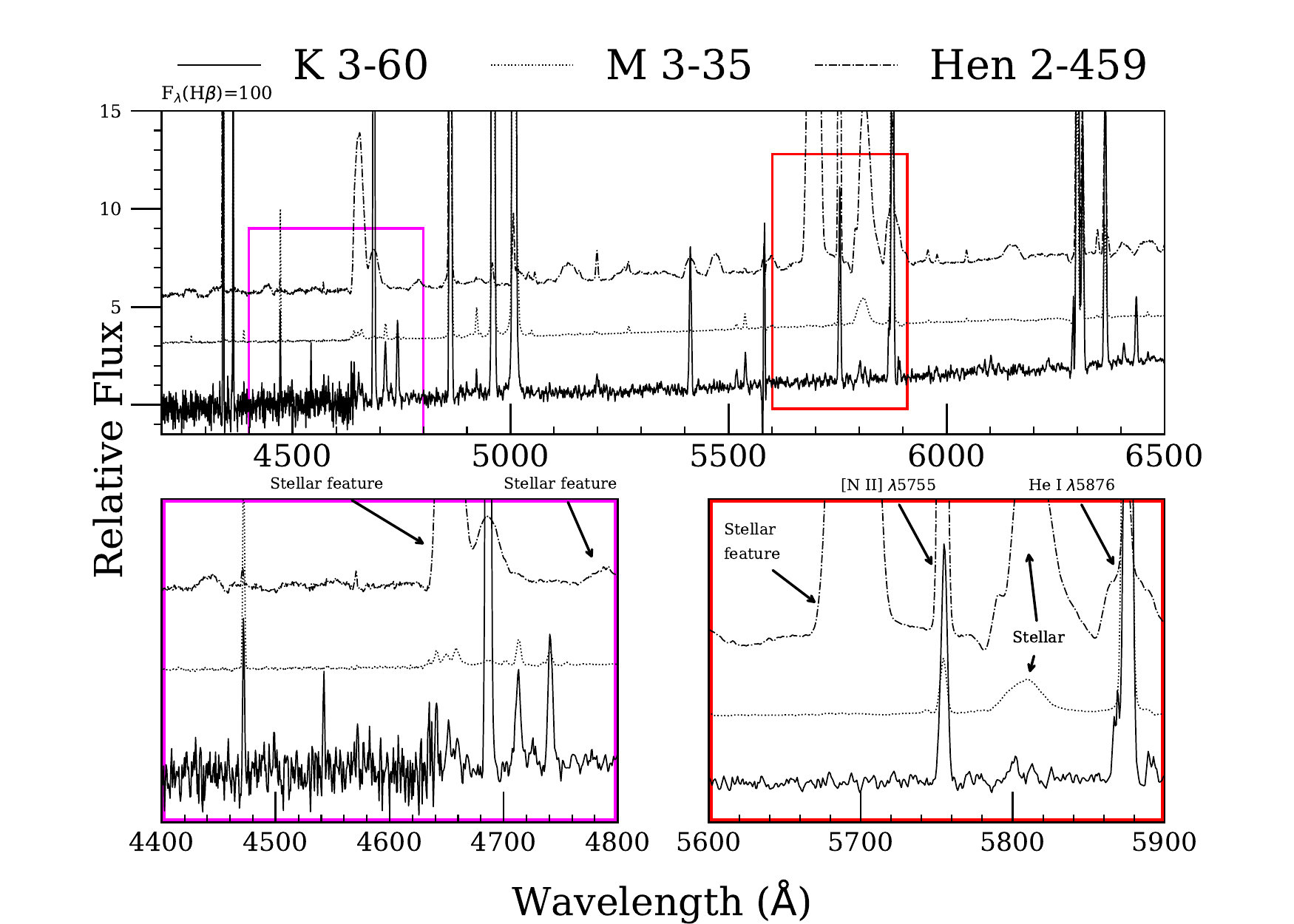}
    \caption{Selected excerpts from the spectra of K 3-60, Hen 2-459 ([WC9] central star with stellar emission features), and M 3-35 (whose central star displays weak stellar emission features). The spectra are scaled to F(H$\beta$) = 100 and vertically shifted to enhance visibility. The bottom two panels zoom in on regions of special interest and illustrate how the presence of stellar emission lines interferes with analysis of the nebular spectrum for objects with emission line stars (see Table \ref{tab:targs}).}
    \label{fig:spectra}
\end{figure}

\subsection{Determination of Physical Conditions and Abundances} \label{subsec:temdens}
We use PyNeb\footnote{\code{http://research.iac.es/proyecto/PyNeb/}} \citep{Pyneb}, a widely-used Python-based nebular analysis package, to determine the physical conditions and ionic and elemental  abundances of our sample. To calculate ionic abundances from collisionally-excited transitions, PyNeb models the ions as $n$-level systems, where $n$ is the number of levels that can be populated via collisional excitation from the ground state at nebular conditions (2 $\leq n \leq$ 34, depending on the ion). It computes equilibrium level populations for specified gas conditions (electron temperature $T_e$  and electron density $n_e$) that are used to calculate the emissivities of various lines emitted by the ion in question. We present the atomic data used in our calculations in Table \ref{tab:atomicdata}.

The energy level structures of certain ions make them favorable for tracing $T_e$  or $n_e$ through intensity ratios of multiple emission lines from the same ion that serve as diagnostics for these parameters. We determine these physical conditions with PyNeb’s \code{getCrossTemDen} function, which takes as input the intensity ratios of a $T_e$ diagnostic and an $n_e$ diagnostic and simultaneously solves for both parameters. When possible, we solve for $T_e$ and $n_e$ using diagnostic ratios from both low-ionization and high-ionization species (Table \ref{tab:tempdens}). Uncertainties for each solution are based on Monte-Carlo simulations that perform 1500 calculations of $T_e$ and $n_e$, allowing the line intensities of the diagnostic ratios to vary within a Gaussian distribution with width equivalent to the line intensity uncertainty. We present the adopted values of $T_e$ and $n_e$ used in the calculation of ionic abundances in Table \ref{tab:tempdens}.
\renewcommand{\arraystretch}{1.3}
\begin{deluxetable*}{r|c|cc|cc}
\tablenum{2}
\tablecaption{Extinction Coefficients, Electron Temperatures, 
and Electron Densities\label{tab:tempdens}}
\tablewidth{0pt}
\tablehead{Object & c(H$\beta$) & Diagnostic & $T_e$  & Diagnostic & $n_e$ \\
\hspace{.2cm} & & & (K) & & (cm$^{-3}$) }
\startdata
Hen 2-459 & 2.1 & \fSIII & 9170$^{+340}_{-460}$ & \fSII & 45830$^{+14250}_{-9860}$ \\
\hline
\multirow{1}{*}{K 3-17} & \multirow{1}{*}{4.2} & \fNII & 11580$^{+140}_{-140}$  & \fSII & 5530$^{+800}_{-640}$ \\
\hline
\multirow{2}{*}{K 3-55} & \multirow{2}{*}{3.5} & \fNII & 11230$^{+210}_{-220}$ & \fSII & 2830$^{+720}_{-810}$ \\
\hspace{.2cm} & & \fOIII & 9810$^{+490}_{-610}$ & \fSII & 2690$^{+1070}_{-680}$ \\
\hline
\multirow{2}{*}{K 3-60} & \multirow{2}{*}{2.1} & \fNII & 13090$^{+240}_{-250}$ & \fClIII & 7980$^{+1380}_{-1270}$ \\
\hspace{.2cm} & & \fOIII & 11940$^{+130}_{-160}$ & \fClIII & 7690$^{+1260}_{-1220}$ \\
\hline
\multirow{2}{*}{K 3-62} & \multirow{2}{*}{2.6} & \fNII & 12150$^{+430}_{-480}$ & \fClIII & 9680$^{+3430}_{-2830}$ \\
\hspace{.2cm} & & \fOIII & 9380$^{+160}_{-180}$ & \fClIII & 9040$^{+3280}_{-2800}$ \\
\hline
\multirow{1}{*}{M 2-43} & \multirow{1}{*}{2.4} &  \fOIII & 10310$^{+390}_{-660}$ & \fClIII & 76880$^{+62760}_{-32690}$ \\
\hline
\multirow{1}{*}{M 3-35} & \multirow{1}{*}{2.2} & \fOIII & 11300$^{+140}_{-160}$ & \fClIII & 19950$^{+2450}_{-2060}$ \\
\hline
\multirow{1}{*}{M 4-18} & \multirow{1}{*}{0.7} & \fNII & 8160$^{+210}_{-280}$ & \fOII, \fSII$^a$ & 9430$^{+4890}_{-2950}$ \\
\enddata
\tablecomments{$T_e$ values are from \fSIII $\lambda 6312/\lambda 9069$, \fNII $\lambda 5755/(\lambda 6548 + \lambda 6584)$, and \fOIII $\lambda 4363/(\lambda 4595 + \lambda 5007)$ diagnostic ratios.  $n_e$ values are from \fSII $\lambda 6731/\lambda 6716$, \fClIII $\lambda 5538/\lambda 5518$, and \fOII $\lambda 3726/\lambda 3729$. For objects with two $T_e$, $n_e$ values, the first applies to the low-ionization zone and the second to the high-ionization zone (see Section \ref{subsec:temdens}).\\
$^a$We adopt the average $n_e$ from these two diagnostic ratios.}
\end{deluxetable*}

Next, we determine ionic abundances from selected lines, excluding the temperature-sensitive auroral lines, using PyNeb. Where possible, we assume a two-zone ionization model for each nebula. For lines from ions with ionization potentials (IPs) \textless~39 eV, we adopt $T_e$ and $n_e$ values determined from the \fNII and \fSII diagnostic ratios respectively. However, in high-density nebulae where $n_e$ $\geq$ 10,000 cm$^{-3}$, the saturation limit of the \fSII $\lambda$6731/6716 diagnostic \citep{Osterbrock2006}, we use \fClIII to determine $n_e$. For lines from ions with IPs $\geq$ 39 eV, we use the \fOIII and \fClIII diagnostic ratios respectively to determine $T_e$ and $n_e$. For Hen 2-459, K 3-17, M 2-43, M 3-35, and M 4-18, we are limited in which diagnostic ratios we can use and thus assume a one-zone ionization model. The \fOIII $\lambda$4363 line is not detected in Hen 2-459 and M 4-18 due to the small abundance of O$^{++}$, and it is not seen in K 3-17 due to the high extinction at this wavelength. Due to the \fNII diagnostic lines returning implausibly large $T_e$ values ($T_e$ \textgreater 20,000K), we are limited to using the \fSIII $T_e$ diagnostic for Hen 2-459.  For M 4-18, we obtain disparate $n_e$ values from the \fOII and \fSII density diagnostics (4160 and 14700 cm$^{-3}$, respectively) and thus choose to adopt the average of the two.  For M 2-43, we are limited to the \fOIII $T_e$  and \fClIII $n_e$ diagnostics as other combinations of diagnostic ratios did not produce plausible values. The ionic abundances are presented in Table \ref{tab:ionabunds}, where the uncertainties were again propagated using Monte Carlo simulations.

Finally, we use PyNeb to compute elemental abundances. When all strongly populated ionic states of an element are observed, one can simply add the ionic abundances to determine an elemental value. However, in many cases, not all significant ions are observed, requiring us to apply ICFs to account for unobserved ions. We adopt the ICFs of \citet{Delgado2014} for all elements except K, Fe, Kr, and Xe. For Fe, we report two abundance values, based on Eq. 2 and Eqs. 3-4 of \citet{Rodriguez2005}. The former ICF was derived from photoionization models, while the latter was determined empirically from observations; together they constrain the range of possible gas-phase Fe abundances \citep{Delgado2015}. We adopt the ICFs for K and Kr from \citet{Amayo2020} and SPD15, respectively. An ICF schema for Xe is not currently available, so we adopt Eq. 7 from SPD15 for Se/Se$^{++}$ due to the similar ionization potential ranges of Xe$^{++}$ and Se$^{++}$. The elemental abundances are presented in Table \ref{tab:elabunds}. We determine uncertainties on the elemental abundances through our Monte Carlo simulation approach, including uncertainties from the ionic abundances and ICFs.

\section{Chemical Abundance Results}\label{sec:chemresults}
\subsection{Overview}\label{sec:overview}
We derive abundances for up to 11 elements per nebula: He, N, O, Ne, S, Cl, Ar, K, Fe, Kr, and Xe. Of these, the $\alpha$-elements O, Ne, S, and Ar are largely unaffected by internal nucleosynthesis in low- and intermediate-mass stars of near solar metallicity \citep[e.g.,][]{Henry2012, Karakas2016}, and Cl, although an odd-numbered element, appears to track the $\alpha$ species \citep{Delgado2014}. These elements should have similar abundances relative to solar for a given PN and are  indicators of the initial composition of the progenitor star. The O and Cl abundances in our sample range from about half solar to solar, while M 3-35 is an outlier with Cl less than one third solar. We also see the widely recognized pattern that S abundances determined for PNe tend to be somewhat lower (by a few tenths of a dex) than those for the other elements in this group. The cause of this effect, sometimes called the “sulfur anomaly” \citep{Henry2004}, is not currently understood \citep{Henry2012, Shingles2013}. The Ar abundances for most of our sample are near solar, except for the two lowest excitation PNe, Hen 2-549 and M 4-18, where the unobserved species Ar$^{+}$ may be more abundant than accounted for by the adopted ICF.

Nebular N abundances have been interpreted as constraining the progenitor mass of a PN. In particular, high enhancements over solar N/O ratios are predicted for intermediate-mass (\textgreater 4~M$_\odot$) AGB stars due to hot bottom burning, where convection mixes material from the H-burning shell into the H-He intershell, enabling proton-captures that produce N from primary C (e.g., \citealp{Karakas2014}). However, there is evidence that N enhancement can also occur in lower-mass progenitors by processes that are not fully understood (e.g., \citealp{Nollett2003}). Our derived N abundances have larger uncertainties than those of some other elements because the only ion we are able to measure is N$^{+}$, although the majority of N atoms will reside in higher ions for high-excitation PNe. In this regard, it is notable that the four objects in our sample for which we derive N abundances of $\geq$ 3 times solar are all objects with moderate to high excitation (K 3-17, K 3-55, K 3-60, and K 3-62). We are using an ICF that assumes N/O = N$^{+}$/O$^{+}$, which may be inaccurate for high-excitation nebulae. Observations in other wavelength regions (UV or far-infrared) would be necessary to directly observe higher ions and more accurately determine the N abundances in these objects.

Fe (and other highly refractory species) cannot be used as indicators of the general metallicity of a PN. These elements became tied up in the solid phase (dust grains) during the late AGB phase of the progenitor star’s life and consequently are depleted out of the nebular gas. The gas-phase abundance of Fe is dominated by this effect and shows depletions of order factors of 10 to 100 \citep{Delgado2015, Delgado2016}.  In our sample, we find [Fe/H] values consistent with that range, although actual values are uncertain due to the fact that we observe lines of only low ions and the ICF formulae are uncertain. 

Finally, optical spectra include emission lines of Kr and Xe, which can be produced by the s-process in AGB stars (\citealp{Pequignot1994, Sharpee2007, GarciaRojas2015}, SPD15, \citealp{Otsuka2020}). Five of the PNe display at least one of the following lines: \fKrIII $\lambda$6827, \fKrIV $\lambda$5346, and $\lambda$5868. Each of the objects in which Kr is detected are enriched in this element relative to O by at least a factor of two, ranging from [Kr/O] = 0.34 dex for M 3-35 and up to 1.5 dex for K 3-60. We also detect \fXeIII $\lambda$10210 in three objects, all of which are likely enriched in Xe. 

\subsection{Planetary Nebulae without Stellar Emission Features}
Four objects in our sample have central stars that are classified as H-rich and do not display stellar emission lines (Table \ref{tab:targs}). Two of these, K 3-17 and K 3-55, have the highest extinctions in our sample. They are the most spatially extended (Table \ref{tab:targs}) and have relatively low $n_e$ values (Table \ref{tab:tempdens}).  Both are high-excitation PNe, for which the ideal $T_e$ diagnostic is the \fOIII line ratio. However, while we measure \fOIII $\lambda$4363 in K 3-55, we do not detect it in K 3-17 due to its high extinction. Both objects appear to have elevated N abundances, although this may be an artifact of their high excitation and an uncertain ICF (see Section \ref{sec:overview}). In both PNe, the O, Cl, and Ar abundances are approximately solar, while S may be somewhat subsolar.  K 3-55 displays Ne lines from which we determine a solar Ne abundance. Previous studies of K 3-17 \citep{Kaler1996} and K 3-55 \citep{Kaler1993} reported H, N, and O abundances in agreement with our findings. We detect K in the spectrum of K 3-17 and determine that [K/H]= $-$0.70 $\pm$ 0.24. Finally, we note a marginal detection of \fKrIII $\lambda$6827 in K 3-17, from which we determine [Kr/O]~= 0.06 $\pm$ 0.17, which is consistent with the lower limit of [Kr/O] \textgreater~0.05 of SPD15.

K 3-60 is a compact, moderate density, high excitation nebula. Infrared spectroscopic observations have shown that it is highly enriched in s-process products, including Kr, Te, and Xe (SPD15, \citealp[][]{Dinerstein2022} and in preparation). Like K 3-17 and K 3-55, it also appears to be enhanced in N and has nearly solar abundances of O and Ar but possibly (modestly) subsolar Cl. We see lines of Ne and determine an abundance of half solar. We also detect K in K 3-60 and find that it is about half solar, suggesting that potassium is not significantly  depleted in this PN. \citet{Aller1987} found He, N, O, Ne, S, and Ar abundances in good agreement with our values. We detect \fKrIV $\lambda$5868, yielding a highly enhanced Kr abundance of [Kr/O] = 1.50 $\pm$ 0.37, consistent within the uncertainties with the value of [Kr/O] = 1.19 $\pm$ 0.75 found by SPD15. We also observed K 3-62, a moderately extended, moderate-excitation nebula for which no previous optical abundance studies are available in the literature, to our knowledge. Similar to K 3-60, K 3-62 shows roughly solar O, Ne, and Ar, but slightly subsolar Cl and K. The Kr abundance, determined from three lines (\fKrIII $\lambda$6827, \fKrIV $\lambda$5346, and \fKrIV $\lambda$5868) is enhanced relative to solar, but not as strongly as for K 3-60. SPD15 were only able to estimate a lower limit of roughly solar Kr for K 3-62 due to the lack of optical data.

\subsection{Planetary Nebulae with Emission-line Central Stars}\label{sec:emission-line}
The other four PNe in our sample display broad stellar emission lines. The presence of such stellar wind lines complicates analysis of the associated nebular spectra. For example, the fluxes of some permitted \NII and \OII lines used to estimate recombination contributions to $T_e$ diagnostics (Section \ref{subsec:temdens}) were not measurable due to blending with stellar features. In the case of Hen  2-459, the strong and broad stellar features prevented measurement of lines used in our abundance analysis including several \fFeIII lines, \fClIII $\lambda$5518, and \fKrIII $\lambda$6827.
\renewcommand{\arraystretch}{1.0}
\setlength{\tabcolsep}{6.5pt}
\movetabledown=5.5cm
\begin{rotatetable}
\begin{deluxetable*}{p{.7cm}p{1.0cm}cccccccc}
\tablenum{3}
\tablecaption{Ionic Abundances\label{tab:ionabunds}}
\tablewidth{0pt}
\tablehead{\colhead{Ion} & \colhead{Line} & \colhead{Hen 2-459} & \colhead{K 3-17} & \colhead{K 3-55} & \colhead{K 3-60} & \colhead{K 3-62} & \colhead{M 2-43} & \colhead{M 3-35} & \colhead{M 4-18}}
\startdata
\HeI & 4471 & $(1.07^{+0.17}_{-0.16})\rm e-2$ &  &  & $(6.14^{+0.58}_{-0.60})\rm e-2$ & $(7.60^{+0.45}_{-0.46})\rm e-2$ & $(9.72^{+0.22}_{-0.33})\rm e-2$ & $(9.55^{+0.10}_{-0.10})\rm e-2$ & $(4.75^{+0.11}_{-0.11})\rm e-2$ \\
\HeI & 5876 & $(2.16^{+0.08}_{-0.09})\rm e-2$ & $(6.51^{+0.48}_{-0.45})\rm e-2$ & $(7.99^{+0.16}_{-0.23})\rm e-2$ & $(6.86^{+0.20}_{-0.21})\rm e-2$ & $(1.00^{+0.04}_{-0.04})\rm e-1$ & $(1.04^{+0.04}_{-0.07})\rm e-1$ & $(1.04^{+0.01}_{-0.01})\rm e-1$ & $(4.72^{+0.22}_{-0.22})\rm e-2$ \\
\HeI & 6678 & $(1.26^{+0.02}_{-0.01})\rm e-2$ & $(5.31^{+0.12}_{-0.13})\rm e-2$ & $(6.40^{+0.24}_{-0.29})\rm e-2$ & $(5.81^{+0.18}_{-0.18})\rm e-2$ & $(8.57^{+0.32}_{-0.36})\rm e-2$ & $(9.75^{+0.39}_{-0.64})\rm e-2$ & $(8.75^{+0.63}_{-0.63})\rm e-2$ & $(6.48^{+0.17}_{-0.18})\rm e-2$ \\
\HeII & 4686 &  & $(4.42^{+0.46}_{-0.46})\rm e-2$ & $(2.43^{+0.22}_{-0.22})\rm e-2$ & $(3.74^{+0.03}_{-0.03})\rm e-2$ & $(3.46^{+0.70}_{-0.70})\rm e-4$ &  &  &  \\
\fNII & 6548 & $(1.11^{+0.19}_{-0.17})\rm e-4$ & $(2.53^{+0.11}_{-0.11})\rm e-5$ & $(4.01^{+0.23}_{-0.22})\rm e-6$ & $(1.23^{+0.05}_{-0.05})\rm e-5$ & $(7.68^{+0.74}_{-0.67})\rm e-6$ & $(1.78^{+0.50}_{-0.65})\rm e-5$ & $(2.56^{+0.10}_{-0.10})\rm e-6$ & $(5.61^{+0.62}_{-0.56})\rm e-5$ \\
\fNII & 6584 & $(1.12^{+0.19}_{-0.18})\rm e-4$ & $(2.61^{+0.09}_{-0.08})\rm e-5$ & $(4.45^{+0.22}_{-0.22})\rm e-5$ & $(1.32^{+0.06}_{-0.06})\rm e-5$ & $(8.36^{+0.85}_{-0.77})\rm e-6$ & $(1.91^{+0.53}_{-0.69})\rm e-5$ & $(2.74^{+0.11}_{-0.10})\rm e-6$ & $(6.23^{+0.68}_{-0.61})\rm e-5$ \\
\fOII & 3726 & $(9.52^{+3.19}_{-3.24})\rm e-5$ &  & $(3.31^{+0.55}_{-0.58})\rm e-5$ & $(1.41^{+0.17}_{-0.16})\rm e-5$ & $(1.41^{+0.31}_{-0.33})\rm e-5$ & $(1.24^{+0.60}_{-0.87})\rm e-4$ & $(1.47^{+0.14}_{-0.15})\rm e-5$ & $(2.46^{+0.59}_{-0.74})\rm e-4$ \\
\fOII & 3729 & $(8.34^{+2.90}_{-2.90})\rm e-5$ &  & $(3.92^{+1.06}_{-1.13})\rm e-5$ & $(1.09^{+0.26}_{-0.25})\rm e-5$ & $(1.60^{+0.43}_{-0.47})\rm e-5$ & $(1.11^{+0.60}_{-0.82})\rm e-4$ & $(1.33^{+0.16}_{-0.18})\rm e-5$ & $(2.90^{+0.85}_{-1.12})\rm e-4$ \\
\fOII & 7320.0 & $(3.87^{+0.66}_{-0.57})\rm e-4$ & $(1.71^{+0.31}_{-0.24})\rm e-4$ & $(8.39^{+1.38}_{-2.11})\rm e-5$ & \multirow{2}{*}{$(2.49^{+0.63}_{-0.65})\rm e-5$} & $(2.24^{+0.55}_{-0.69})\rm e-5$ & $(2.78^{+2.26}_{-4.59})\rm e-4$ & \multirow{2}{*}{$(4.16^{+0.27}_{-0.28})\rm e-5$} & $(3.77^{+1.44}_{-2.35})\rm e-4$ \\
\fOII & 7330.0 & $(6.05^{+1.12}_{-1.35})\rm e-4$ & $(2.76^{+0.51}_{-0.37})\rm e-4$ & $(1.02^{+0.15}_{-0.24})\rm e-4$ &  & $(3.32^{+0.8}_{-0.91})\rm e-5$ & $(2.96^{+1.61}_{-3.83})\rm e-4$ &  & $(5.86^{+1.89}_{-3.64})\rm e-4$ \\
\fOIII & 4959 & $(1.62^{+0.34}_{-0.26})\rm e-6$ & $(3.80^{+0.15}_{-0.15})\rm e-4$ & $(6.06^{+1.49}_{-1.15})\rm e-4$ & $(2.66^{+0.11}_{-0.09})\rm e-4$ & $(4.13^{+0.29}_{-0.27})\rm e-4$ & $(1.57^{+0.38}_{-0.24})\rm e-4$ & $(2.35^{+0.11}_{-0.09})\rm e-4$ &  \\
\fOIII & 5007 & $(1.78^{+0.38}_{-0.28})\rm e-6$ & $(4.04^{+0.16}_{-0.15})\rm e-4$ & $(6.38^{+1.50}_{-1.20})\rm e-4$ & $(2.82^{+0.11}_{-0.09})\rm e-4$ & $(4.31^{+0.31}_{-0.27})\rm e-4$ & $(1.63^{+0.41}_{-0.26})\rm e-4$ & $(2.16^{+0.10}_{-0.10})\rm e-4$ & $(1.25^{+0.30}_{-0.27})\rm e-7$ \\
\fNeIII & 3869 &  &  & $(9.71^{+3.87}_{-3.45})\rm e-5$ & $(4.43^{+0.22}_{-0.19})\rm e-5$ & $(9.15^{+0.79}_{-0.70})\rm e-5$ & $(4.79^{+1.83}_{-1.41})\rm e-7$ & $(5.23^{+0.28}_{-0.25})\rm e-5$ &  \\
\fNeIII & 3967 &  &  &  & $(4.29^{+0.22}_{-0.19})\rm e-5$ & $(8.05^{+0.74}_{-0.64})\rm e-5$ &  & $(5.11^{+0.28}_{-0.24})\rm e-5$ &  \\
\fSII & 6716 & $(4.81^{+1.18}_{-1.49})\rm e-6$ & $(4.22^{+0.35}_{-0.41})\rm e-7$ & $(1.27^{+0.20}_{-0.27})\rm e-6$ & $(4.83^{+0.65}_{-0.67})\rm e-7$ & $(1.94^{+0.44}_{-0.53})\rm e-7$ & $(4.80^{+2.14}_{-3.23})\rm e-7$ & $(2.07^{+0.22}_{-0.25})\rm e-7$ & $(3.83^{+0.94}_{-1.53})\rm e-6$ \\
\fSII & 6731 & $(4.81^{+1.17}_{-1.42})\rm e-6$ & $(4.22^{+0.26}_{-0.30})\rm e-7$ & $(1.27^{+0.13}_{-0.15})\rm e-6$ & $(5.05^{+0.52}_{-0.54})\rm e-7$ & $(1.97^{+0.38}_{-0.44})\rm e-7$ & $(5.22^{+2.23}_{-3.76})\rm e-7$ & $(2.00^{+0.24}_{-0.24})\rm e-7$ & $(4.00^{+0.88}_{-1.26})\rm e-6$ \\
\fSIII & 6312 & $(6.05^{+1.48}_{-1.12})\rm e-6$ & $(2.24^{+0.15}_{-0.15})\rm e-6$ & $(4.12^{+0.37}_{-0.37})\rm e-6$ & $(2.41^{+0.16}_{-0.14})\rm e-6$ & $(1.07^{+0.16}_{-0.14})\rm e-6$ & $(3.96^{+1.16}_{-0.72})\rm e-6$ & $(2.10^{+0.12}_{-0.10})\rm e-6$ &  \\
\fSIII & 9069 & $(6.05^{+0.66}_{-0.52})\rm e-6$ & $(2.70^{+0.06}_{-0.06})\rm e-6$ & $(2.44^{+0.09}_{-0.09})\rm e-6$ & $(2.41^{+0.08}_{-0.07})\rm e-6$ & $(2.13^{+0.15}_{-0.14})\rm e-6$ & $(4.66^{+0.65}_{-0.56})\rm e-6$ & $(2.00^{+0.27}_{-0.29})\rm e-6$ & $(5.38^{+0.42}_{-0.32})\rm e-7$ \\
\fSIII & 9531 & $(3.80^{+0.42}_{-0.31})\rm e-6$ & $(2.84^{+0.06}_{-0.06})\rm e-6$ & $(3.79^{+0.14}_{-0.13})\rm e-6$ & $(1.72^{+0.18}_{-0.17})\rm e-6$ &  & $(4.30^{+0.62}_{-0.47})\rm e-6$ & $(1.79^{+0.11}_{-0.10})\rm e-6$ &  \\
\fClII & 8579 & $(1.09^{+0.13}_{-0.09})\rm e-7$ & $(9.09^{+0.31}_{-0.31})\rm e-9$ & $(1.90^{+0.08}_{-0.08})\rm e-8$ & $(8.55^{+0.29}_{-0.28})\rm e-9$ & $(4.41^{+0.39}_{-0.35})\rm e-9$ & $(1.19^{+0.17}_{-0.13})\rm e-8$ & $(5.36^{+0.54}_{-0.53})\rm e-9$ & $(9.54^{+0.84}_{-0.61})\rm e-8$ \\
\fClII & 9124 & $(1.30^{+0.16}_{-0.12})\rm e-7$ & $(9.08^{+0.78}_{-0.76})\rm e-9$ & $(2.15^{+0.16}_{-0.15})\rm e-8$ & $(9.00^{+0.62}_{-0.61})\rm e-9$ & $(7.33^{+0.85}_{-0.85})\rm e-9$ & $(1.20^{+0.19}_{-0.16})\rm e-8$ &  & $(9.21^{+1.48}_{-1.38})\rm e-8$ \\
\fClIII & 5518 &  & $(7.76^{+2.40}_{-2.62})\rm e-8$ &  & $(4.57^{+0.45}_{-0.45})\rm e-8$ & $(7.79^{+1.58}_{-1.69})\rm e-8$ & $(5.16^{+2.00}_{-3.14})\rm e-8$ & $(3.36^{+0.29}_{-0.30})\rm e-8$ & $(5.20^{+1.47}_{-1.76})\rm e-8$ \\
\fClIII & 5538 & $(3.58^{+1.14}_{-1.08})\rm e-8$ &  & $(1.98^{+0.43}_{-0.36})\rm e-7$ & $(4.57^{+0.23}_{-0.22})\rm e-8$ & $(7.79^{+0.73}_{-0.68})\rm e-8$ & $(7.07^{+2.33}_{-2.57})\rm e-8$ & $(3.36^{+0.15}_{-0.14})\rm e-8$ & $(8.75^{+1.49}_{-1.39})\rm e-8$ \\
\fClIV & 7531 &  & $(9.51^{+0.25}_{-0.23})\rm e-8$ & $(6.54^{+1.13}_{-0.97})\rm e-8$ & $(6.21^{+0.47}_{-0.46})\rm e-8$ & $(2.81^{+0.86}_{-0.87})\rm e-8$ & $(1.26^{+0.40}_{-0.36})\rm e-9$ & $(1.24^{+0.05}_{-0.05})\rm e-8$ &  \\
\fClIV & 8046 &  & $(6.48^{+0.17}_{-0.16})\rm e-8$ & $(3.58^{+0.55}_{-0.43})\rm e-8$ & $(4.39^{+0.12}_{-0.10})\rm e-8$ & $(1.21^{+0.15}_{-0.15})\rm e-8$ &  & $(8.25^{+0.26}_{-0.22})\rm e-9$ &  \\
\fArIII & 7136 & $(7.77^{+1.02}_{-0.75})\rm e-8$ & $(1.69^{+0.05}_{-0.05})\rm e-6$ & $(2.04^{+0.33}_{-0.26})\rm e-6$ & $(1.18^{+0.04}_{-0.03})\rm e-6$ & $(1.60^{+0.08}_{-0.07})\rm e-6$ & $(1.55^{+0.24}_{-0.14})\rm e-6$ & $(7.35^{+0.27}_{-0.25})\rm e-7$ & $(3.57^{+1.07}_{-1.06})\rm e-8$ \\
\fArIII & 7751 & $(6.22^{+1.93}_{-1.77})\rm e-8$ & $(1.53^{+0.04}_{-0.04})\rm e-6$ & $(1.86^{+0.31}_{-0.24})\rm e-6$ & $(1.22^{+0.03}_{-0.03})\rm e-6$ & $(1.58^{+0.07}_{-0.07})\rm e-6$ & $(1.65^{+0.26}_{-0.14})\rm e-6$ & $(8.13^{+0.25}_{-0.22})\rm e-7$ &  \\
\enddata
\tablecomments{Abundances presented as $\rm \frac{N(X)}{N(H)}$}
\end{deluxetable*}
\end{rotatetable}
\renewcommand{\arraystretch}{1.0}
\setlength{\tabcolsep}{6.5pt}
\movetabledown=5.5cm
\begin{rotatetable}
\begin{deluxetable*}{p{.7cm}p{1.0cm}cccccccc}
\tablenum{3}
\tablecaption{Ionic Abundances (continued)\label{tab:ionabunds}}
\tablewidth{0pt}
\tablehead{\colhead{Ion} & \colhead{Line} & \colhead{Hen 2-459} & \colhead{K 3-17} & \colhead{K 3-55} & \colhead{K 3-60} & \colhead{K 3-62} & \colhead{M 2-43} & \colhead{M 3-35} & \colhead{M 4-18}}
\startdata
\fArIV & 4711 &  &  &  & $(7.84^{+0.86}_{-0.83})\rm e-7$ &  &  &  &  \\
\fArIV & 4740 &  & $(1.53^{+0.14}_{-0.14})\rm e-6$ &  & $(7.20^{+0.58}_{-0.58})\rm e-7$ & $(2.49^{+0.24}_{-0.26})\rm e-7$ &  & $(8.95^{+0.41}_{-0.37})\rm e-8$ &  \\
\fArV & 6435 &  & $(3.81^{+0.15}_{-0.14})\rm e-7$ & $(6.64^{+2.22}_{-1.99})\rm e-8$ & $(2.47^{+0.26}_{-0.29})\rm e-7$ &  &  &  &  \\
\fArV & 7005 &  & $(3.44^{+0.11}_{-0.10})\rm e-7$ & $(8.45^{+1.57}_{-1.23})\rm e-8$ & $(2.28^{+0.08}_{-0.06})\rm e-7$ &  &  &  &  \\
\fKIV & 6102 &  & $(8.39^{+0.81}_{-0.76})\rm e-9$ &  & $(1.30^{+0.14}_{-0.15})\rm e-8$ & $(3.32^{+1.23}_{-1.26})\rm e-9$ & $(1.73^{+0.62}_{-0.51})\rm e-10$ & $(5.23^{+1.13}_{-1.10})\rm e-10$ &  \\
\fFeII & 8617 & $(3.09^{+0.36}_{-0.27})\rm e-7$ & $(3.07^{+0.20}_{-0.20})\rm e-8$ &  & $(7.32^{+1.81}_{-1.92})\rm e-9$ &  & $(6.34^{+0.89}_{-0.68})\rm e-10$ & $(4.29^{+0.45}_{-0.45})\rm e-8$ &  \\
\fFeIII & 4702 &  &  &  &  &  &  & $(1.31^{+0.15}_{-0.15})\rm e-7$ & $(1.03^{+0.23}_{-0.23})\rm e-6$ \\
\fFeIII & 4734 &  &  &  &  &  & $(3.76^{+1.59}_{-3.06})\rm e-7$ &  &  \\
\fFeIII & 4755 &  &  &  &  &  & $(4.45^{+1.18}_{-0.84})\rm e-7$ & $(1.60^{+0.22}_{-0.21})\rm e-7$ & $(3.83^{+0.74}_{-0.67})\rm e-6$ \\
\fFeIII & 5270 &  &  &  &  &  & $(4.18^{+0.92}_{-0.52})\rm e-7$ & $(1.46^{+0.07}_{-0.06})\rm e-7$ & $(1.71^{+0.34}_{-0.31})\rm e-6$ \\
\fKrIII & 6827 &  & $(3.38^{+0.84}_{-0.83})\rm e-9$ &  &  & $(1.07^{+0.58}_{-0.59})\rm e-9$ & $(7.30^{+1.83}_{-1.37})\rm e-9$ & $(8.61^{+0.81}_{-0.81})\rm e-10$ &  \\
\fKrIV & 5346 &  &  &  &  & $(7.71^{+1.36}_{-1.34})\rm e-9$ &  &  &  \\
\fKrIV & 5868 &  &  &  & $(1.44^{+0.12}_{-0.12})\rm e-8$ & $(8.27^{+1.31}_{-1.30})\rm e-9$ &  &  &  \\
\fXeIII & 10210 & $(3.28^{+0.83}_{-0.77})\rm e-10$ &  &  &  &  & $(5.85^{+0.85}_{-0.73})\rm e-10$ &  & $(2.71^{+0.47}_{-0.47})\rm e-9$ \\
\enddata
\tablecomments{Abundances presented as $\rm \frac{N(X)}{N(H)}$}
\end{deluxetable*}
\end{rotatetable}
\renewcommand{\arraystretch}{1.3}
\movetabledown=6.5cm
\begin{rotatetable}
\begin{deluxetable*}{rllllllll|l}
\tablenum{4}
\tablecaption{Elemental Abundances\label{tab:elabunds}}
\tablewidth{0pt}
\tablehead{\colhead{Element} & \colhead{Hen 2-459} & \colhead{K 3-17} & \colhead{K 3-55} & \colhead{K 3-60} & \colhead{K 3-62} & \colhead{M 2-43} & \colhead{M 3-35} & \colhead{M 4-18} & {$\odot$ }} 
\startdata
$\rm \frac{N(He)}{N(H)}$ & $>$0.019 & 0.107$\pm$0.006 & 0.101$\pm$0.003 & 0.103$\pm$0.004 & 0.095$\pm$0.004 & 0.103$\pm$0.004 & 0.100$\pm$0.004 & $>$0.052 & 0.08 \\
\hline
$\rm \frac{N(N)}{N(H)}$ & (1.13$\pm$0.44)e-4 & (1.99$\pm$0.37)e-4 & (8.31$\pm$2.43)e-4 & (3.58$\pm$0.85)e-4 & (2.14$\pm$0.58)e-4 & (3.25$\pm$1.33)e-5 & (4.34$\pm$0.56)e-5 & (6.09$\pm$2.48)e-5 & 6.76e-5 \\
$[$N/H] & 0.22$\pm$0.22 & 0.47$\pm$0.09 & 1.09$\pm$0.15 & 0.72$\pm$0.12 & 0.50$\pm$0.14 & -0.32$\pm$0.23 & -0.19$\pm$0.06 & -0.05$\pm$0.23 & 7.83 \\
\hline
$\rm \frac{N(O)}{N(H)}$ &  (3.44$\pm$1.01)e-4 & (6.86$\pm$1.16)e-4 & (7.91$\pm$1.63)e-4 & (3.88$\pm$0.65)e-4 & (4.44$\pm$0.39)e-4 & (3.82$\pm$0.98)e-4 & (2.36$\pm$0.15)e-4 & (2.83$\pm$0.91)e-4 & 4.90e-4 \\
$[$O/H] & -0.15$\pm$0.15 & 0.15$\pm$0.08 & 0.21$\pm$0.10 & -0.10$\pm$0.08 & -0.04$\pm$0.04 & -0.11$\pm$0.13 & -0.32$\pm$0.03 & -0.24$\pm$0.17 & 8.69 \\
\hline
$\rm \frac{N(Ne)}{N(H)}$ &  &  & (1.14$\pm$0.39)e-4 & (5.85$\pm$1.21)e-5 & (1.01$\pm$0.20)e-4 & (2.47$\pm$0.45)e-6 & (8.00$\pm$1.18)e-5 &  & 1.1e-4 \\
$[$Ne/H] &  &  & 0.00$\pm$0.18 & -0.29$\pm$0.10 & -0.06$\pm$0.10 & -1.67$\pm$0.09$^{a}$ & -0.16$\pm$0.07 &  & 8.06 \\
\hline
$\rm \frac{N(S)}{N(H)}$ & (9.52$\pm$1.23)e-6 & (5.51$\pm$1.33)e-6 & (9.94$\pm$3.08)e-6 & (6.19$\pm$1.70)e-6 & (4.51$\pm$1.38)e-6 & (4.91$\pm$0.72)e-6 & (2.06$\pm$0.18)e-6 & (4.48$\pm$0.75)e-6 & 1.41e-5 \\
$[$S/H] & -0.17$\pm$0.06 & -0.41$\pm$0.12 & -0.15$\pm$0.16 & -0.36$\pm$0.14 & -0.50$\pm$0.16 & -0.46$\pm$0.07 & -0.84$\pm$0.04 & -0.50$\pm$0.08 & 7.12 \\
\hline
$\rm \frac{N(Cl)}{N(H)}$ & (1.51$\pm$0.58)e-7 & (1.57$\pm$0.35)e-7 & (2.59$\pm$0.62)e-7 & (1.01$\pm$0.22)e-7 & (1.01$\pm$0.24)e-7 & (8.09$\pm$1.67)e-8 & (4.85$\pm$0.22)e-8 & (1.74$\pm$0.69)e-7 & 1.70e-7 \\
$[$Cl/H] & -0.05$\pm$0.21 & -0.03$\pm$0.11 & 0.18$\pm$0.12 & -0.22$\pm$0.11 & -0.23$\pm$0.12 & -0.32$\pm$0.10 & -0.54$\pm$0.02 & 0.01$\pm$0.22 & 5.23 \\
\hline
$\rm \frac{N(Ar)}{N(H)}$ &  (3.75$\pm$1.85)e-7 & (3.00$\pm$1.28)e-6 & (3.64$\pm$1.64)e-6 & (2.49$\pm$1.03)e-6 & (2.59$\pm$1.09)e-6 & (1.69$\pm$0.78)e-6 & (1.16$\pm$0.48)e-6 & (2.04$\pm$1.02)e-7 & 2.40e-6 \\
$[$Ar/H] & -0.81$\pm$0.30 & 0.10$\pm$0.24 & 0.18$\pm$0.26 & 0.02$\pm$0.23 & 0.03$\pm$0.24 & -0.15$\pm$0.27 & -0.32$\pm$0.24 & -1.07$\pm$0.30 & 6.38 \\
\hline
$\rm \frac{N(K)}{N(H)}$ &  & (2.43$\pm$1.02)e-8 &  & (6.19$\pm$2.62)e-8 & (7.22$\pm$3.61)e-8 &  & (1.48$\pm$0.64)e-8 &  & 1.20e-7 \\
$[$K/H] &  & -0.70$\pm$0.24 &  & -0.29$\pm$0.24 & -0.22$\pm$0.3 &  & -0.91 $\pm$ 0.25 &  & 5.07 \\
\hline
\multirow{2}{*}{$\rm \frac{N(Fe)}{N(H)}$} & \multirow{2}{*}{(3.09$\pm$2.16)e-7$^{R4}$} &  &  &  &  & (6.61$\pm$2.14)e-7$^{R2}$ & (1.66$\pm$0.28)e-6$^{R2}$ & (3.98$\pm$1.10)e-6$^{R2}$ & \multirow{2}{*}{2.82e-5} \\
  &   &  &  &  &  & (7.25$\pm$2.35)e-7$^{R3}$  & (5.25$\pm$0.68)e-7$^{R3}$ &  (2.40$\pm$0.66)e-6$^{R4}$  &   \\
\multirow{2}{*}{$[$Fe/H]} & \multirow{2}{*}{-1.96$\pm$0.52$^{R4}$} &  &  &  &  & -1.63$\pm$0.17$^{R2}$ & -1.23$\pm$0.08$^{R2}$ & -0.85$\pm$0.14$^{R2}$ & \multirow{2}{*}{7.46} \\
  &   &  &  &  &  & -1.59$\pm$0.17$^{R3}$  & -1.73$\pm$0.06$^{R3}$ &  -1.07$\pm$0.14$^{R4}$  &   \\
  \hline
$\rm \frac{N(Kr)}{N(H)}$ &  & (1.06$\pm$0.30)e-8 &  & (3.33$\pm$1.87)e-8 & (3.71$\pm$1.53)e-9 & (9.96$\pm$3.54)e-9 & (2.67$\pm$1.14)e-9 &  & 1.32e-9 \\
$[$Kr/H] &  & 0.90$\pm$0.15 &  & 1.40$\pm$0.36 & 0.45$\pm$0.24 & 0.88$\pm$0.15 & 0.02$\pm$0.07 &  & 3.12 \\
\hline
$\rm \frac{N(Xe)}{N(H)}$ & (4.31$\pm$2.40)e-9 &  &  &  &  & (8.42$\pm$8.46)e-10 &  & (4.13$\pm$2.35)e-8 & 1.66e-10 \\
$[$Xe/H] &  1.41$\pm$0.36 &  &  &  &  & 0.81$\pm$0.41 &  & 2.40$\pm$0.61 & 2.22 \\
\enddata
\tablecomments{Elemental abundances presented both linearly (first row) and logarithmically relative to solar (second row, where [X/H]= $\rm log_{10}\frac{N(X)}{N(H)} - log_{10}\frac{N(X)_\odot}{N(H)_\odot}$), except for He, which is only reported linearly. He abundances for Hen 2-459 and M 4-18 are lower limits since they do not include contributions from He$^0$, which is significant in such low-excitation PNe. Fe abundances are reported separately for the R2 and R3 + R4 ICFs of \citet{Rodriguez2005}. Solar abundances, presented in linear (top) and logarithmic (log $\epsilon(\odot)$, bottom) forms, are photospheric values from \citet{Asplund2021}, except for Cl, which is the meteoritic value from \citet{Palme2014}.\\
$^{a}$Based on a marginal Ne detection.}
\end{deluxetable*}
\end{rotatetable}
Hen 2-459 and M 4-18 both have cool, very late-type [WC] central stars \citep{Acker2003}. In Table \ref{tab:elabunds}, we report their He abundances based on \HeI recombination lines as lower limits, since much of the He in these low-excitation nebulae is in the form of unobservable neutral He. Also due to the low excitation, we did not detect the intrinsically weak \fOIII $\lambda$4363 line in either of these objects. Both objects show slightly subsolar O abundances but Cl abundances consistent with the solar value within measurement uncertainties. Ar appears to be deficient in Hen 2-459 and M 4-18, although---as in the case of He---there are significant reservoirs of Ar in the unobserved neutral and singly-ionized states that are not fully accounted for by the Ar ICF. \citet{Girard2007} found He, N, Ne, Ar, and S abundances similar to ours for Hen 2-459 but significantly lower O and Cl, while \citet{Sabbadin1980} and \citet{demarco1999} found O, N, and S abundances that agree well with our determinations. \fXeIII $\lambda$10210 is present in the spectra of Hen 2-459 and M 4-18 , and Xe appears to be enriched in both objects by large factors (Table \ref{tab:elabunds}).

The Wolf-Rayet central star of M 2-43 is type [WC7-8], earlier than those of Hen 2-459 and M 4-18. Like the latter PNe, M 2-43 is compact and high-density, but the nebular excitation is higher due to the hotter central star. The abundances of alpha species O and Cl, and also N, are about half solar. The very low apparent Ne abundance, however, is likely spurious since it is based on a marginal detection of \fNeIII $\lambda$3869. \citet{Girard2007} reported a higher Ne abundance than ours, but similar abundances for He, N, O, S, Ar, and Cl. \fKIV 6102 is detected in this PN, but we are unable to determine an abundance due to the lack of \fArIV{}, a required input of the \citealt{Amayo2020} ICF. M 2-43 is the only object in our sample to display lines of both Kr and Xe. We find that Kr/O and Xe/O are both approximately ten times solar, where the former is in good agreement with the Kr/O ratio found by SPD15 from near-infrared observations of \fKrIII.

We present the first detection of stellar emission lines in the spectrum of M 3-35 (see Figure \ref{fig:spectra}), indicating that it possesses a weak emission-line (WEL) or [WC] central star. The nebula is moderately extended, with a complex, multipolar morphology \citep{Hsia2014}. It appears to be the lowest metallicity object in our sample, according to the abundances of O, Cl, and Ar. \citet{Barker1978} reported similar abundances for these elements. In this PN, the [K/O] abundance ratio indicates that potassium is depleted by a factor of four.  We detect lines of Fe that indicate a (typical) depletion of about a factor of 30 \citep[e.g.,][]{Delgado2015, Delgado2016}, depending on the choice of ICF (see Table \ref{tab:elabunds} footnote). Similarly, Hen 2-459, M 2-43, and M 4-18 display Fe depletion factors of 10-100.  M 3-35 shows the \fKrIII $\lambda$6827 and \fKrIV $\lambda$5346 lines and appears to have at most a modest enrichment in Kr, about a factor of two relative to O. 

\section{Spatiokinematic Analysis and Population Memberships}\label{sec:kin}
To supplement the interpretation of our chemical abundance results, we utilize the kinematics of our PNe to estimate their probabilities of belonging to the thin disk, thick disk, and halo of the Milky Way.  We make use of \Gaia{} eDR3 \citep{GaiaeDR3} positions and proper motions, which are available for all of our targets except K 3-17 and K 3-60, and radial velocities from our LRS2 data. We adopt distances from \citet{Bailer2021}, who used a Bayesian approach to estimate distances to \Gaia{} sources. They adopt an exponentially-decreasing distance prior that, when applied to the parallax and its uncertainties, reports more accurate distance estimates than those derived from direct parallax inversion. The distance prior treats the line-of-sight stellar density as exponentially decreasing with increasing distance from the Galactic center. We determine Galactic space velocities (UVW) for our sample using the \code{SkyCoord} module of Astropy \citep{Astropy}, with positions, proper motions, radial velocities, and distances as inputs. We transform the V velocity components to the LSR frame by subtracting 235 km/s, the circular velocity at the local standard of rest \citep{Schonrich2010}, and determine uncertainties of the positions and velocities through Monte Carlo error propagation. 

\begin{figure}
    \centering

        \centering
        \includegraphics[width=.7\linewidth]{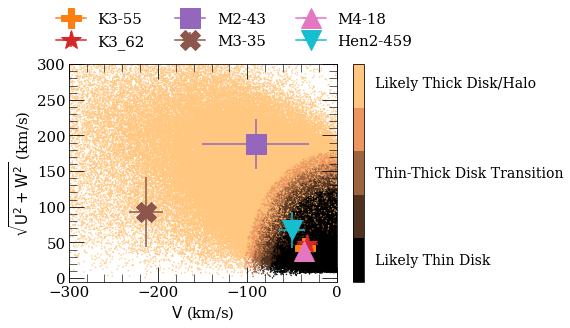} 
        \includegraphics[width=.7\linewidth]{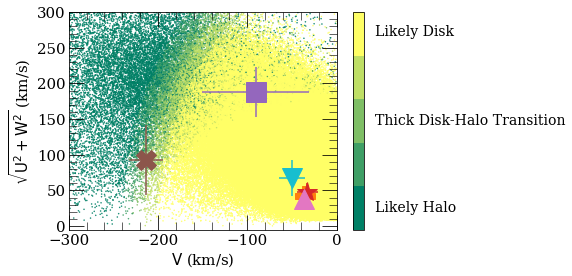} 
    \caption{Toomre diagrams depicting the kinematics of the PNe in our sample with \Gaia{} data. The background of each panel is based on a subsample of \Gaia{} stars, colored by thick disk to thin disk membership probability for the top panel and halo to thick disk membership probability for the bottom panel. \label{fig:toomre}
    }
\end{figure}

Next, we compute the relative probabilities of population memberships based on kinematics alone for each PN with \Gaia{} data. We refer readers to Section 2 of \citet{Bensby2003} for a description of the procedures for computing relative probabilities of membership. We follow the approach of \citet{Carrillo2020} in interpreting the results of our membership analysis.  Thin disk members are defined as having D/TD \textgreater~10, which indicates that the PN has a more than 10 times greater likelihood of belonging to the thin disk (D) than to the thick disk (TD). Likely halo members (H) have H/TD \textgreater~10, and objects in the thick disk-halo and thin disk-thick disk transition zones have 0.5 \textless~TD/H \textless~2 and 0.5 \textless~TD/D \textless~2, respectively. We then compare the Galactic space velocities and membership probabilities of each PN with the space velocities of a subsample of \Gaia{} eDR3 stars from \citet{Marchetti2021} and their associated membership probabilities that we calculate and define using the same methods. We present the Galactocentric positions and Galactic space velocities of our sample in Table \ref{tab:xyzuvw} and the kinematic analysis in Figure \ref{fig:toomre}. We compare the Galactocentric positions of our sample to the Galactic spiral arm model of \citet{Reid2014} to estimate their present locations in the Milky Way. 

Hen 2-459, K 3-55, K 3-62, and M 4-18 show strong kinematic signs of thin disk membership, with 160~\textless~D/TD \textless~191. The first three of these are located at low Galactic Z values, less than 100 pc from the Galactic plane. Hen 2-459 and K 3-55 lie along the Local arm of the Galaxy, while K 3-62 appears to lie along the Perseus arm. M 4-18 is the second most distant object in our sample. It lies 5 kpc away from the Sun in the Outer arm and is nearly 1 kpc above the Galactic plane. 
M 2-43’s kinematics suggest that it is a thick disk member. Of all our PNe, this object is closest to the Galactic Center, located approximately halfway between the Sun and the Galactic center at a distance of 4 kpc. This object appears to lie along the Scutum arm and is approximately 350 pc above the Galactic plane. Finally, M 3-35 is the most distant object in our sample, more than 6 kpc from the Sun in the Perseus arm. While its kinematics place it in the transition region between the halo and the thick disk (Figure \ref{fig:toomre}), its small vertical distance Z from the Galactic Plane (Table \ref{tab:xyzuvw}) and only moderately subsolar abundances (Table \ref{tab:elabunds}) suggest that it belongs to the thick disk rather than the halo.

We can now use the combined information from the kinematics and chemistry of the six PNe with both kinds of information to assess their populations of origin. Hen 2-459, K 3-55, K 3-62, and M 4-18 possess roughly solar metallicities, consistent with their kinematic signatures of thin disk membership. M 3-35, with the lowest metallicity of our sample, shows chemical and kinematic signs of thick disk membership. M 2-43’s kinematic properties indicate that it is a member of the thick disk,  and its low Cl abundance seems to support this. However, its O abundance is solar within  the uncertainties, making its population membership less clear than that of M 3-35. K3-17 and K3-60 lack \Gaia{} astrometric data but have metallicities indicative of thin disk membership.

\renewcommand{\arraystretch}{1.3}
\begin{deluxetable*}{rrrrrrrrr}
\tablenum{5}
\tablecaption{Radial Velocities, Distances, Galactic Positions and Velocities\label{tab:xyzuvw}}
\tablewidth{0pt}
\tablehead{\colhead{Object} & \colhead{RV} & \colhead{Distance} & \colhead{X} & \colhead{Y} & \colhead{Z} & \colhead{U} & \colhead{V} & \colhead{W}\\
\colhead{ } & \colhead{(km s$^{-1}$)} & \colhead{(kpc)} & \colhead{(kpc)} & \colhead{(kpc)} & \colhead{(kpc)} & \colhead{(km s$^{-1}$)} & \colhead{(km s$^{-1}$)} & \colhead{(km s$^{-1}$)}}
\startdata
Hen 2-459 & -36$\pm$11 & 1.67$^{+0.82}_{-0.58}$ &  7.51$\pm$0.21 & 1.55$\pm$0.53 & -0.06$\pm$0.03 & 65.3$\pm$23.1 & -50.4$\pm$14.7 & -17.5$\pm$9.6 \\
K 3-55 & -33$\pm$12 & 1.86$^{+0.2}_{-0.17}$ & 7.48$\pm$0.06 & 1.74$\pm$0.16 & 0.02$\pm$0.00 & 41.4$\pm$5.5 & -35.2$\pm$11.3 & -3.4$\pm$1.1 \\
K 3-62 & -46.5$\pm$12 & 3.33$^{+1.98}_{-1.18}$ &  8.42$\pm$0.11 & 3.31$\pm$1.16 & 0.07$\pm$0.02 & 47.9$\pm$11.8 & -32.8$\pm$11.9 & 1.5$\pm$5.0 \\
M 2-43 & 107$\pm$10 & 4.13$^{+1.48}_{-1.69}$ &  4.47$\pm$1.47 & 1.91$\pm$0.77 & 0.32$\pm$0.12 & 187.8$\pm$33.7 & -90.5$\pm$60.0 & -0.2$\pm$8.7 \\
M 3-35 & -189$\pm$11 & 6.24$^{+2.09}_{-2.1}$ &  6.16$\pm$0.66 & 5.92$\pm$1.98 & -0.24$\pm$0.09 & 78.5$\pm$43.0 & -213.1$\pm$18.9 & -49.8$\pm$23.7 \\
M 4-18 & -47$\pm$10 & 5.23$^{+0.67}_{-0.44}$ &  12.46$\pm$0.36 & 2.84$\pm$0.24 & 0.72$\pm$0.06 & 37.6$\pm$9.9 & -36.5$\pm$6.7 & 1.4$\pm$1.6 \\
\enddata
\tablecomments{Radial velocities, distances from \citealt{Bailer2021}, and Galactocentric XYZ positions and UVW space velocities for PNe in our sample with \Gaia{} data.}
\end{deluxetable*}
\clearpage
\section{Summary}\label{sec:sum}
\begin{enumerate}
    \item We present new determinations of the physical conditions and ionic and elemental abundances in up to 11 elements for eight highly-extincted Milky Way PNe that lack high-quality optical data in the previous literature.
    \item We report the discovery of broad stellar emission features in M 3-35's spectrum indicative of a WEL or [WC] central star.
    \item The O abundances of the sample range from half-solar to solar. A few PNe appear to show lower abundances for the heavier $\alpha$-species tracers Ar and Cl, although this may be due to uncertainties in our ionization correction procedure. Similarly, uncertainties in correcting for unobserved N ions may be responsible for the large reported abundances of N in several PNe.  In four PNe in which Fe was detected, we find Fe depletion factors of 10-100, while K is depleted by factors of 2-7 in four objects.
    \item  For PNe with \Gaia{} astrometric data, three-dimensional Galactocentric positions and space velocities are presented and plotted in kinematic phase space diagrams that distinguish between the Milky Way’s thin disk, thick disk, and halo populations.
    \item Four PNe show chemical and kinematic signs of thin disk membership, and two show chemical and kinematic signs of thick disk membership.  The final two lack \Gaia{} data but possess chemical compositions suggestive of thin disk membership.
    \item Four PNe are enriched in Kr and three in Xe, both elements that can be synthesized by the s-process. Our newly determined nebular parameters and abundances of the light elements can be used to supplement future spectroscopic studies of these nebulae, including observations of infrared emission lines from trans-iron species made by AGB stars that contribute to the chemical evolution of the Galaxy.
\end{enumerate}

\section*{Acknowledgements}
We thank Roger Wesson for advice on adapting ALFA to best suit the analysis of our data and Zachary G. Maas for interesting and useful discussions pertaining to Cl.  This work was supported by NSF grant AST 17-15332.  The Low Resolution Spectrograph 2 (LRS2) was developed and funded by the University of Texas at Austin McDonald Observatory and Department of Astronomy and by the Pennsylvania State University. We thank the Leibniz-Institut f\"{u}r Astrophysik Potsdam (AIP) and the Institut f\"{u}r Astrophysik G\"{o}ttingen (IAG) for their contributions to the construction of the integral field units.  The Hobby-Eberly Telescope (HET) is a joint project of the University of Texas at Austin, the Pennsylvania State University, Ludwig-Maximilians-Universit\"{a}t München, and Georg-August-Universit\"{a}t G\"{o}ttingen. The HET is named in honor of former Texas Lieutenant-Governor William P. Hobby and Robert Eberly, a principal benefactor from the Pennsylvania State University.

\bibliography{lrs2}
\bibliographystyle{aasjournal}

\appendix

\renewcommand{\arraystretch}{1.0}
\movetabledown=4.5cm
\begin{rotatetable}
\begin{deluxetable*}{rlcccccccc}
\tablenum{A1}
\tablecaption{Observed Fluxes \label{tab:fluxlist}}
\tablewidth{0pt}
\tablehead{\multicolumn{10}{c}{Fluxes}\\\colhead{Ion} & \colhead{Wave} & \colhead{Hen 2-459} & \colhead{K 3-17} & \colhead{K 3-55} & \colhead{K 3-60} & \colhead{K 3-62} & \colhead{M 2-43} & \colhead{M 3-35} & \colhead{M 4-18}}
\startdata
\HeI & 4471 & 0.64 $\pm$ 0.10 &  &  & 2.12 $\pm$ 0.20 & 2.32 $\pm$ 0.13 & 3.14 $\pm$ 0.05 & 3.06 $\pm$ 0.03 & 2.15 $\pm$ 0.60 \\
\HeI & 5876 & 20.31 $\pm$ 0.75 & 107.47 $\pm$ 7.53 & 77.94 $\pm$ 0.69 & 40.89 $\pm$ 0.86 & 73.91 $\pm$ 0.57 & 65.07 $\pm$ 0.67 & 56.57 $\pm$ 0.29 & 13.00 $\pm$ 0.61 \\
\HeI & 6678 & 7.82 $\pm$ 0.43 & 87.14 $\pm$ 1.88 & 49.56 $\pm$ 1.77 & 18.46 $\pm$ 0.44 & 38.39 $\pm$ 0.76 & 34.33 $\pm$ 0.69 & 25.34 $\pm$ 1.83 & 6.01 $\pm$ 0.69 \\
\HeII & 4686 &  & 34.44 $\pm$ 3.76 & 19.96 $\pm$ 1.77 & 36.60 $\pm$ 0.29 & 0.33 $\pm$ 0.07$^*$ &  &  &  \\
\fNII & 5755 & 102.38 $\pm$ 0.08 & 34.46 $\pm$ 0.53 & 24.73 $\pm$ 0.88 & 11.07 $\pm$ 0.19 & 6.57 $\pm$ 0.11 & 13.46 $\pm$ 0.30 & 2.53 $\pm$ 0.05 & 3.25 $\pm$ 0.09 \\
\fNII & 6548 & 968.25 $\pm$ 1.59 & 1187.65 $\pm$ 37.64 & 135.16 $\pm$ 3.53 & 303.12 $\pm$ 1.72 & 150.04 $\pm$ 1.07 & 125.90 $\pm$ 2.99 & 25.56 $\pm$ 0.24 & 129.10 $\pm$ 0.61 \\
\fNII & 6584 & 2960.32 $\pm$ 1.59 & 3802.00 $\pm$ 37.64 & 4611.31 $\pm$ 2.65 & 980.84 $\pm$ 2.86 & 496.20 $\pm$ 10.72 & 409.69 $\pm$ 2.99 & 82.71 $\pm$ 0.24 & 425.76 $\pm$ 1.53 \\
\fOII & 3726 & 7.54 $\pm$ 0.19 &  & 6.17 $\pm$ 0.88 & 8.59 $\pm$ 0.50 & 4.42 $\pm$ 0.21 & 5.14 $\pm$ 0.09 & 2.85 $\pm$ 0.09 & 65.57 $\pm$ 1.22 \\
\fOII & 3729 & 2.60 $\pm$ 0.18 &  & 4.47 $\pm$ 0.88 & 3.17 $\pm$ 0.61 & 2.31 $\pm$ 0.17 & 1.78 $\pm$ 0.09 & 1.08 $\pm$ 0.08 & 34.95 $\pm$ 1.21 \\
\fOII & 7320 & 330.95 $\pm$ 0.79 & 714.85 $\pm$ 14.18 & 253.89 $\pm$ 2.53 & 110.07 $\pm$ 1.13 & 66.34 $\pm$ 0.91 & 362.44 $\pm$ 3.88 & 57.56 $\pm$ 0.84 & 18.46 $\pm$ 0.03 \\
\fOII & 7330 & 304.76 $\pm$ 0.79 & 682.10 $\pm$ 7.44 & 177.83 $\pm$ 1.69 & 80.33 $\pm$ 1.11 & 57.28 $\pm$ 0.65 & 226.97 $\pm$ 4.51 & 45.74 $\pm$ 0.51 & 16.63 $\pm$ 0.11 \\
\fOIII & 4363 &  &  & 4.92 $\pm$ 1.01 & 8.88 $\pm$ 0.25 & 2.55 $\pm$ 0.07 & 2.33 $\pm$ 0.04 & 4.98 $\pm$ 0.04 &  \\
\fOIII & 4959 & 2.40 $\pm$ 0.16 & 825 $\pm$ 13 & 689 $\pm$ 6 & 557 $\pm$ 6 & 401 $\pm$ 5 & 191 $\pm$ 2 & 388 $\pm$ 3 &  \\
\fOIII & 5007 & 8.18 $\pm$ 0.48 & 2880 $\pm$ 30 & 2330 $\pm$ 20 & 1820 $\pm$ 20 & 139 $\pm$ 14 & 618 $\pm$ 6 & 1104 $\pm$ 24 & 0.24 $\pm$ 0.05 \\
\fNeIII & 3869 &  &  & 10.87 $\pm$ 2.65 & 24.19 $\pm$ 0.40 & 14.86 $\pm$ 0.12 & 0.14 $\pm$ 0.03 & 21.97 $\pm$ 0.12 &  \\
\fNeIII & 3967 &  &  &  & 7.98 $\pm$ 0.20 & 4.57 $\pm$ 0.12 &  & 7.32 $\pm$ 0.08 &  \\
\fSII & 4069 & 12.13 $\pm$ 0.26 &  &  & 1.04 $\pm$ 0.19 & 0.44 $\pm$ 0.09$^*$ & 0.63 $\pm$ 0.03 & 0.58 $\pm$ 0.02 & 4.61 $\pm$ 2.29 \\
\fSII & 4076 & 4.23 $\pm$ 0.17 &  &  &  &  & 0.27 $\pm$ 0.02 & 0.26 $\pm$ 0.02 & 3.64 $\pm$ 0.23 \\
\fSII & 6716 & 54.21 $\pm$ 0.79 & 148.97 $\pm$ 2.28 & 302.12 $\pm$ 17.67 & 29.74 $\pm$ 1.72 & 12.97 $\pm$ 0.11 & 3.05 $\pm$ 0.03 & 4.41 $\pm$ 0.24 & 23.65 $\pm$ 0.46 \\
\fSII & 6731 & 123.02 $\pm$ 2.38 & 282.33 $\pm$ 5.11 & 493.82 $\pm$ 26.50 & 61.48 $\pm$ 1.72 & 26.90 $\pm$ 0.11 & 7.66 $\pm$ 0.21 & 9.31 $\pm$ 0.72 & 50.36 $\pm$ 0.15 \\
\fSIII & 6312 & 17.94 $\pm$ 0.24 & 43.48 $\pm$ 1.88 & 36.22 $\pm$ 1.77 & 16.01 $\pm$ 0.09 & 7.22 $\pm$ 0.05 & 13.34 $\pm$ 0.09 & 7.716 $\pm$ 0.002 &  \\
\fSIII & 9069 & 1254 $\pm$ 1 & 6983 $\pm$ 8 & 2783 $\pm$ 1 & 858 $\pm$ 9 & 842 $\pm$ 1 & 969 $\pm$ 1 & 352 $\pm$ 48 & 9.67 $\pm$ 0.08 \\
\fSIII & 9531 & 2270 $\pm$ 2 & 24657 $\pm$ 8 & 13693 $\pm$ 1 & 1764 $\pm$ 172 &  & 2616 $\pm$ 2 & 909 $\pm$ 48 &  \\
\fClII & 8579 & 57.94 $\pm$ 0.05 & 71.43 $\pm$ 1.84 & 48.22 $\pm$ 1.02 & 5.25 $\pm$ 0.06 & 4.34 $\pm$ 0.22 & 5.95 $\pm$ 0.26 & 2.46 $\pm$ 0.24 & 4.00 $\pm$ 0.09 \\
\fClII & 9124 & 22.38 $\pm$ 1.03 & 28.38 $\pm$ 2.35 & 20.23 $\pm$ 1.33 & 1.80 $\pm$ 0.11 & 2.45 $\pm$ 0.23 & 1.99 $\pm$ 0.19 &  & 1.09 $\pm$ 0.15 \\
\fClIII & 5518 &  & 3.52 $\pm$ 1.13 &  & 0.90 $\pm$ 0.06 & 0.83 $\pm$ 0.11 & 0.20 $\pm$ 0.03 & 0.34 $\pm$ 0.01 & 0.21 $\pm$ 0.04 \\
\fClIII & 5538 & 0.77 $\pm$ 0.19 &  & 5.63 $\pm$ 0.27 & 1.52 $\pm$ 0.06 & 1.56 $\pm$ 0.11 & 1.05 $\pm$ 0.06 & 0.88 $\pm$ 0.01 & 0.68 $\pm$ 0.07 \\
\fClIV & 7531 &  & 56.28 $\pm$ 0.19 & 10.87 $\pm$ 0.88 & 4.06 $\pm$ 0.29 & 1.79 $\pm$ 0.54 & 0.08 $\pm$ 0.02$^*$ & 0.71 $\pm$ 0.02 &  \\
\fClIV & 8046 &  & 227.74 $\pm$ 1.88 & 32.16 $\pm$ 0.35 & 13.15 $\pm$ 0.03 & 3.73 $\pm$ 0.43 &  & 2.18 $\pm$ 0.02 &  \\
\fArIII & 5192 &  &  &  &  &  & 0.19 $\pm$ 0.01 & 0.18 $\pm$ 0.02 &  \\
\fArIII & 7136 & 10.71 $\pm$ 0.16 & 1727.27 $\pm$ 12.65 & 641.52 $\pm$ 6.12 & 166.89 $\pm$ 2.11 & 205.66 $\pm$ 2.37 & 197.46 $\pm$ 2.06 & 90.14 $\pm$ 1.85 & 0.61 $\pm$ 0.18 \\
\fArIII & 7751 & 2.93 $\pm$ 0.79 & 760.40 $\pm$ 3.76 & 249.12 $\pm$ 8.83 & 59.19 $\pm$ 0.09 & 74.80 $\pm$ 0.21 & 74.46 $\pm$ 0.18 & 34.24 $\pm$ 0.12 &  \\
\fArIV & 4711 &  &  &  & 3.29 $\pm$ 0.29 &  &  & 0.846 $\pm$ 0.002 &  \\
\fArIV & 4740 &  & 6.83 $\pm$ 0.56 &  & 4.35 $\pm$ 0.29 & 0.67 $\pm$ 0.03 &  & 0.490 $\pm$ 0.005 &  \\
\enddata
\tablecomments{Table continued on next page.}
\end{deluxetable*}
\end{rotatetable}
\renewcommand{\arraystretch}{1.0}
\movetabledown=4.5cm
\begin{rotatetable}
\begin{deluxetable*}{rlcccccccc}
\tablenum{A1}
\tablecaption{Observed Fluxes \label{tab:fluxlist}}
\tablewidth{0pt}
\tablehead{\multicolumn{10}{c}{Fluxes}\\\colhead{Ion} & \colhead{Wave} & \colhead{Hen 2-459} & \colhead{K 3-17} & \colhead{K 3-55} & \colhead{K 3-60} & \colhead{K 3-62} & \colhead{M 2-43} & \colhead{M 3-35} & \colhead{M 4-18}}
\startdata
\fArIV & 7263 &  &  & 3.84 $\pm$ 0.27 & 2.06 $\pm$ 0.11 & 0.37 $\pm$ 0.12 &  & 0.12 $\pm$ 0.03 &  \\
\fArV & 6435 &  & 27.48 $\pm$ 0.56 & 1.63 $\pm$ 0.44 & 4.03 $\pm$ 0.46 &  &  &  &  \\
\fArV & 7005 &  & 147.94 $\pm$ 1.13 & 10.69 $\pm$ 0.27 & 15.01 $\pm$ 0.09 &  &  &  &  \\
\fKIV & 6102 &  & 2.15 $\pm$ 0.19 &  & 0.97 $\pm$ 0.10 & 0.17 $\pm$ 0.06 & 0.010 $\pm$ 0.003 & 0.033 $\pm$ 0.007 &  \\
\fFeII & 8617 & 7.79 $\pm$ 0.05 & 12.16 $\pm$ 0.72 &  & 0.23 $\pm$ 0.06 &  & 0.014 $\pm$ 0.0003 & 0.97 $\pm$ 0.10 &  \\
\fFeIII & 4702 &  &  &  &  &  &  & 0.12 $\pm$ 0.01 & 0.47 $\pm$ 0.07$^*$ \\
\fFeIII & 4734 &  &  &  &  &  & 0.17 $\pm$ 0.04 &  &  \\
\fFeIII & 4755 &  &  &  &  &  & 0.17 $\pm$ 0.02 & 0.08 $\pm$ 0.01 & 1.01 $\pm$ 0.13$^*$ \\
\fFeIII & 5270 &  &  &  &  &  & 0.88 $\pm$ 0.04 & 0.379 $\pm$ 0.009 & 1.52 $\pm$ 0.24 \\
\fKrIII & 6827 &  & 2.29 $\pm$ 0.56 &  &  & 0.19 $\pm$ 0.11 & 0.74 $\pm$ 0.12 & 0.085 $\pm$ 0.007 &  \\
\fKrIV & 5346 &  &  &  &  & 0.19 $\pm$ 0.03 &  & 0.011 $\pm$ 0.001&  \\
\fKrIV & 5868 &  &  &  & 1.53 $\pm$ 0.11 & 0.58 $\pm$ 0.09$^*$ &  &  &  \\
\fXeIII & 10210 & 1.02 $\pm$ 0.24 &  &  &  &  & 1.79 $\pm$ 0.18 &  & 0.46 $\pm$ 0.08 \\
\enddata
\tablecomments{Observed line fluxes and associated uncertainties. We correct line fluxes for interstellar dust extinction (Table \ref{tab:linelist}) before proceeding with nebular parameter and abundance analysis. Entries marked with $^*$ denote a marginal detection.}
\end{deluxetable*}
\end{rotatetable}

\renewcommand{\arraystretch}{1.0}
\movetabledown=4.5cm
\begin{rotatetable}
\begin{deluxetable*}{rlcccccccc}
\tablenum{A2}
\tablecaption{Intensities \label{tab:linelist}}
\tablewidth{0pt}
\tablehead{\multicolumn{10}{c}{Fluxes}\\\colhead{Ion} & \colhead{Wave} & \colhead{Hen 2-459} & \colhead{K 3-17} & \colhead{K 3-55} & \colhead{K 3-60} & \colhead{K 3-62} & \colhead{M 2-43} & \colhead{M 3-35} & \colhead{M 4-18}}
\startdata
Ion & Wave & Hen 2-459 & K 3-17 & K 3-55 & K 3-60 & K 3-62 & M 2-43 & M 3-35 & M 4-18 \\
\hline
\HeI & 4471 & 0.56 $\pm$ 0.09 &  &  & 3.38 $\pm$ 0.32 & 4.12 $\pm$ 0.24 & 5.19 $\pm$ 0.08 & 5.14 $\pm$ 0.05 & 2.45 $\pm$ 0.06 \\
\HeI & 5876 & 3.40 $\pm$ 0.13 & 10.49 $\pm$ 0.73 & 12.17 $\pm$ 0.11 & 12.21 $\pm$ 0.26 & 17.34 $\pm$ 0.13 & 17.20 $\pm$ 0.18 & 17.59 $\pm$ 0.09 & 7.13 $\pm$ 0.34 \\
\HeI & 6678 & 0.565 $\pm$ 0.003 & 2.40 $\pm$ 0.05 & 2.74 $\pm$ 0.10 & 2.90 $\pm$ 0.07 & 4.13 $\pm$ 0.08 & 4.49 $\pm$ 0.09 & 4.12 $\pm$ 0.30 & 2.76 $\pm$ 0.07 \\
\HeII & 4686 &  & 52.43 $\pm$ 5.73 & 29.62 $\pm$ 2.62 & 44.11 $\pm$ 0.34 & 0.42 $\pm$ 0.09$^*$ &  &  &  \\
\fNII & 5755 & 19.10 $\pm$ 0.01 & 4.17 $\pm$ 0.06 & 4.61 $\pm$ 0.16 & 3.69 $\pm$ 0.06 & 1.76 $\pm$ 0.03 & 4.01 $\pm$ 0.09 & 0.83 $\pm$ 0.02 & 1.84 $\pm$ 0.05 \\
\fNII & 6548 & 94.29 $\pm$ 0.15 & 39.44 $\pm$ 1.25 & 8.72 $\pm$ 0.23 & 52.37 $\pm$ 0.30 & 18.14 $\pm$ 0.13 & 18.27 $\pm$ 0.43 & 4.58 $\pm$ 0.04 & 59.82 $\pm$ 0.28 \\
\fNII & 6584 & 280.75 $\pm$ 0.15 & 119.78 $\pm$ 1.19 & 285.06 $\pm$ 0.16 & 165.00 $\pm$ 0.48 & 58.07 $\pm$ 1.25 & 57.74 $\pm$ 0.42 & 14.41 $\pm$ 0.04 & 195.67 $\pm$ 0.70 \\
\fOII & 3726 & 94.33 $\pm$ 0.15 &  & 8.72 $\pm$ 0.23 & 36.11 $\pm$ 0.20 & 18.74 $\pm$ 0.13 & 16.91 $\pm$ 0.40 & 4.75 $\pm$ 0.04 & 50.26 $\pm$ 0.24 \\
\fOII & 3729 & 280.88 $\pm$ 0.15 &  & 285.06 $\pm$ 0.16 & 113.76 $\pm$ 0.33 & 60.01 $\pm$ 1.30 & 53.45 $\pm$ 0.39 & 14.96 $\pm$ 0.04 & 164.38 $\pm$ 0.59 \\
\fOII & 7320 & 6.65 $\pm$ 0.14 & 1.37 $\pm$ 0.17 & 2.52 $\pm$ 0.07 & 3.79 $\pm$ 0.11 & 1.74 $\pm$ 0.06 & 7.64 $\pm$ 0.32 & 2.41 $\pm$ 0.10 & 7.82 $\pm$ 0.41 \\
\fOII & 7330 & 17.67 $\pm$ 0.05 & 8.07 $\pm$ 0.09 & 4.93 $\pm$ 0.05 & 8.22 $\pm$ 0.11 & 3.67 $\pm$ 0.04 & 18.54 $\pm$ 0.37 & 4.82 $\pm$ 0.05 & 6.56 $\pm$ 0.04 \\
\fOIII & 4363 &  &  & 14.17 $\pm$ 2.91 & 16.14 $\pm$ 0.46 & 5.33 $\pm$ 0.15 & 4.46 $\pm$ 0.07 & 9.59 $\pm$ 0.07 &  \\
\fOIII & 4959 & 1.11 $\pm$ 0.07 & 604 $\pm$ 9 & 561 $\pm$ 5 & 463 $\pm$ 5 & 325 $\pm$ 4 & 155 $\pm$ 2 & 339 $\pm$ 3 &  \\
\fOIII & 5007 & 3.53 $\pm$ 0.21 & 1855 $\pm$ 19 & 1709 $\pm$ 13 & 1417 $\pm$ 14 & 980 $\pm$ 10 & 467 $\pm$ 5 & 904 $\pm$ 20 & 0.18 $\pm$ 0.03 \\
\fNeIII & 3869 &  &  & 82.45 $\pm$ 20.11 & 80.15 $\pm$ 1.32 & 64.25 $\pm$ 0.53 & 0.50 $\pm$ 0.11 & 77.56 $\pm$ 0.44 &  \\
\fNeIII & 3967 &  &  &  & 23.38 $\pm$ 0.57 & 17.02 $\pm$ 0.46 &  & 22.84 $\pm$ 0.25 &  \\
\fSII & 4069 & 32.49 $\pm$ 0.02 &  &  & 26.80 $\pm$ 0.27 & 20.80 $\pm$ 0.03$^*$ & 30.00 $\pm$ 0.03 & 16.94 $\pm$ 2.32 & 2.50 $\pm$ 0.02 \\
\fSII & 4076 & 50.41 $\pm$ 0.04 &  &  &  &  & 68.30 $\pm$ 0.04 & 37.37 $\pm$ 1.98 & 2.752 $\pm$ 0.003 \\
\fSII & 6716 & 4.68 $\pm$ 0.07 & 3.89 $\pm$ 0.06 & 16.00 $\pm$ 0.94 & 4.55 $\pm$ 0.26 & 1.35 $\pm$ 0.01 & 0.393 $\pm$ 0.004 & 0.70 $\pm$ 0.04 & 10.55 $\pm$ 0.20 \\
\fSII & 6731 & 10.50 $\pm$ 0.20 & 7.21 $\pm$ 0.13 & 25.71 $\pm$ 1.38 & 9.30 $\pm$ 0.26 & 2.77 $\pm$ 0.01 & 0.96 $\pm$ 0.03 & 1.46 $\pm$ 0.11 & 22.40 $\pm$ 0.07 \\
\fSIII & 6312 & 2.093 $\pm$ 0.002 & 1.88 $\pm$ 0.09 & 3.03 $\pm$ 0.15 & 3.16 $\pm$ 0.02 & 1.09 $\pm$ 0.01 & 2.36 $\pm$ 0.02 & 1.662 $\pm$ 0.001&  \\
\fSIII & 9069 & 32.48 $\pm$ 0.02 & 16.61 $\pm$ 0.02 & 20.70 $\pm$ 0.01 & 38.87 $\pm$ 0.39 & 20.13 $\pm$ 0.03 & 32.41 $\pm$ 0.03 & 16.31 $\pm$ 2.23 & 2.97 $\pm$ 0.02 \\
\fSIII & 9531 & 50.39 $\pm$ 0.04 & 43.16 $\pm$ 0.01 & 79.22 $\pm$ 0.01 & 68.40 $\pm$ 6.65 &  & 73.78 $\pm$ 0.04 & 35.99 $\pm$ 1.91 &  \\
\fClII & 8579 & 1.814 $\pm$ 0.004& 0.25 $\pm$ 0.01 & 0.49 $\pm$ 0.01 & 0.294 $\pm$ 0.005& 0.13 $\pm$ 0.01 & 0.24 $\pm$ 0.01 & 0.14 $\pm$ 0.01 & 1.30 $\pm$ 0.03 \\
\fClII & 9124 & 0.57 $\pm$ 0.03 & 0.07 $\pm$ 0.01 & 0.15 $\pm$ 0.01 & 0.082 $\pm$ 0.004& 0.06 $\pm$ 0.01 & 0.07 $\pm$ 0.01 &  & 0.33 $\pm$ 0.05 \\
\fClIII & 5518 &  & 0.67 $\pm$ 0.22 &  & 0.38 $\pm$ 0.02 & 0.29 $\pm$ 0.04 & 0.08 $\pm$ 0.01 & 0.15 $\pm$ 0.01 & 0.13 $\pm$ 0.03 \\
\fClIII & 5538 & 0.18 $\pm$ 0.04 &  & 1.48 $\pm$ 0.07 & 0.63 $\pm$ 0.02 & 0.54 $\pm$ 0.04 & 0.40 $\pm$ 0.02 & 0.38 $\pm$ 0.01 & 0.41 $\pm$ 0.04 \\
\fClIV & 7531 &  & 0.534 $\pm$ 0.003& 0.25 $\pm$ 0.02 & 0.37 $\pm$ 0.03 & 0.10 $\pm$ 0.03 & 0.013 $\pm$ 0.004$^*$ & 0.072 $\pm$ 0.003&  \\
\fClIV & 8046 &  & 1.26 $\pm$ 0.01 & 0.48 $\pm$ 0.01 & 0.913 $\pm$ 0.001& 0.15 $\pm$ 0.02 &  & 0.161 $\pm$ 0.005&  \\
\fArIII & 5192 &  &  &  &  &  & 0.11 $\pm$ 0.01 & 0.11 $\pm$ 0.02 &  \\
\fArIII & 7136 & 0.70 $\pm$ 0.01 & 25.85 $\pm$ 0.19 & 21.55 $\pm$ 0.21 & 19.24 $\pm$ 0.24 & 15.23 $\pm$ 0.18 & 18.39 $\pm$ 0.19 & 10.72 $\pm$ 0.22 & 0.25 $\pm$ 0.07 \\
\fArIII & 7751 & 0.13 $\pm$ 0.04 & 5.64 $\pm$ 0.03 & 4.71 $\pm$ 0.17 & 4.78 $\pm$ 0.01 & 3.60 $\pm$ 0.01 & 4.69 $\pm$ 0.01 & 2.84 $\pm$ 0.01 &  \\
\fArIV & 4711 &  &  &  & 3.83 $\pm$ 0.33 &  &  & 1.041 $\pm$ 0.003 &  \\
\fArIV & 4740 &  & 9.01 $\pm$ 0.74 &  & 4.87 $\pm$ 0.32 & 0.78 $\pm$ 0.04 &  & 0.578 $\pm$ 0.006 &  \\
\enddata
\tablecomments{Table continued on next page.}
\end{deluxetable*}
\end{rotatetable}
\renewcommand{\arraystretch}{1.0}
\movetabledown=4.5cm
\begin{rotatetable}
\begin{deluxetable*}{rlcccccccc}
\tablenum{A2}
\tablecaption{Intensities \label{tab:linelist}}
\tablewidth{0pt}
\tablehead{\multicolumn{10}{c}{Fluxes}\\\colhead{Ion} & \colhead{Wave} & \colhead{Hen 2-459} & \colhead{K 3-17} & \colhead{K 3-55} & \colhead{K 3-60} & \colhead{K 3-62} & \colhead{M 2-43} & \colhead{M 3-35} & \colhead{M 4-18}}
\startdata
\fArIV & 7263 &  & 0.303 $\pm$ 0.016 & 0.114 $\pm$ 0.008 & 0.219 $\pm$ 0.011 & 0.025 $\pm$ 0.008 &  & 0.013 $\pm$ 0.004 &  \\
\fArV & 6435 &  & 1.08 $\pm$ 0.02 & 0.12 $\pm$ 0.03 & 0.76 $\pm$ 0.09 &  &  &  &  \\
\fArV & 7005 &  & 2.61 $\pm$ 0.02 & 0.41 $\pm$ 0.01 & 1.88 $\pm$ 0.01 &  &  &  &  \\
\fKIV & 6102 &  & 0.14 $\pm$ 0.01 &  & 0.24 $\pm$ 0.03 & 0.03 $\pm$ 0.014 & 0.002 $\pm$ 0.001 & 0.008 $\pm$ 0.002 &  \\
\fFeII & 8617 & 0.240 $\pm$ 0.002 & 0.041 $\pm$ 0.002 &  & 0.012 $\pm$ 0.003 &  & 0.0010 $\pm$ 0.0001 & 0.053 $\pm$ 0.005 &  \\
\fFeIII & 4702 &  &  &  &  &  &  & 0.15 $\pm$ 0.02 & 0.40 $\pm$ 0.06$^*$ \\
\fFeIII & 4734 &  &  &  &  &  & 0.19 $\pm$ 0.05 &  &  \\
\fFeIII & 4755 &  &  &  &  &  & 0.19 $\pm$ 0.02 & 0.10 $\pm$ 0.01 & 0.83 $\pm$ 0.11$^*$ \\
\fFeIII & 5270 &  &  &  &  &  & 0.46 $\pm$ 0.02 & 0.221 $\pm$ 0.005 & 1.01 $\pm$ 0.16 \\
\fKrIII & 6827 &  & 0.05 $\pm$ 0.01 &  &  & 0.02 $\pm$ 0.01 & 0.09 $\pm$ 0.01 & 0.012 $\pm$ 0.001 &  \\
\fKrIV & 5346 &  &  &  &  & 0.09 $\pm$ 0.01 &  &  &  \\
\fKrIV & 5868 &  &  &  & 0.46 $\pm$ 0.03 & 0.14 $\pm$ 0.02$^*$ &  &  &  \\
\fXeIII & 10210 & 0.019 $\pm$ 0.004 &  &  &  &  & 0.041 $\pm$ 0.004 &  & 0.13 $\pm$ 0.02 \\
\enddata
\tablecomments{Extinction-corrected intensities and associated uncertainties, adopting c(H$\beta$) from Table \ref{tab:tempdens}. The contents of this table are used in our nebular parameter and abundance analysis.  Entries marked with $^*$ denote a marginal detection.}
\end{deluxetable*}
\end{rotatetable}

\renewcommand{\arraystretch}{1.0}
\begin{deluxetable*}{l|l|l}
\tablenum{A3}
\tablecaption{Atomic Data\label{tab:atomicdata}}
\tablehead{Ion & Transition Probabilities& Collision Strengths}
\startdata
N$^{+}$ & \citet{Froese2004} & \citet{Tayal2011} \\
O$^{+}$ & \citet{Froese2004} & \citet{Kisielius2009} \\
O$^{++}$ & \citet{Froese2004} & \citet{Storey2014} \\
  & \citet{Storey2000} &  \\
Ne$^{++}$ & \citet{Galavis1997} & \citet{Mclaughlin2000} \\
S$^{+}$ & \citet{Podobedova2009} & \citet{Tayal2010} \\
S$^{++}$ & \citet{Podobedova2009} & \citet{Tayal1999} \\
Cl$^{+}$ & \citet{Mendoza1983} & \citet{Tayal2004} \\
Cl$^{++}$ & \citet{Mendoza1983b} & \citet{Butler1989} \\
Cl$^{3+}$ & \citet{Kaufman1986} & \citet{Galavis1995} \\
 & \citet{Mendoza1982a} &  \\
 & \citet{Ellis1984} &  \\
Ar$^{++}$ & \citet{Mendoza1983b} & \citet{Galavis1995} \\
 & \citet{Kaufman1986} &  \\
Ar$^{3+}$ & \citet{Mendoza1982b} & \citet{Ramsbottom1997} \\
Ar$^{4+}$ & \citet{Mendoza1982a} & \citet{Galavis1995} \\
 & \citet{Kaufman1986} &  \\
 & \citet{LaJohn1993} &  \\
K$^{3+}$ & \citet{Kaufman1986} & \citet{Galavis1995} \\
 & \citet{Mendoza1983b} &  \\
Fe$^{++}$ & \citet{Quinet1996} & \citet{Zhang1996} \\
 & \citet{Johansson2000} &  \\
Kr$^{++}$ & \citet{Biemont1986a} & \citet{Schoning1997}  \\
Kr$^{3+}$ & \citet{Biemont1986b}  &   \citet{Schoning1997} \\
Xe$^{++}$ & \citet{Biemont1995}  &   \citet{Schoening1998} \\
\hline
\hline
Ion & Recombination Coefficients &\\
\hline
H$^{+}$ & \citet{Storey1995} &\\
He$^{+}$ & \citet{Porter2012} &\\
He$^{++}$ & \citet{Storey1995} &\\
N$^{++}$ & \citet{Fang2011, Fang2013} &\\
 & \citet{Kisielius2002} &  \\
 & \citet{Escalante1992} &  \\
O$^{++}$ & \citet{Storey1994} &\\
 & \citet{Liu1995} &  \\
\enddata
\end{deluxetable*}

\end{document}